\documentclass[smallextended]{svjour3}       
\smartqed 

\usepackage[utf8]{inputenc} 

\usepackage{bbm}
\usepackage{geometry} 
\usepackage{graphicx}
\usepackage{epstopdf}
\usepackage{doi}

\usepackage{amsmath,amsfonts,amssymb}

\newcommand{\field}[1]{\mathbb{#1}}
\newcommand {\R}{\field{R} }
\newcommand {\N}{ {\field{N}} }

\newcommand{\eins}{\mathbbm{1}}
\newcommand {\ol}{\overline }

\newtheorem{prop}{Proposition}[section]
\newtheorem{theo}[prop]{Theorem}
\newtheorem{ddef}[prop]{Definition}
\newtheorem{rem}[prop]{Remark}

\newtheorem{lem}[prop]{Lemma}
\newtheorem{cor}[prop]{Corollary}

\usepackage{color}
\usepackage{colordvi}
\definecolor{magenta}{rgb}{.5,0,.5}
\definecolor{black}{rgb}{1.0,1.0,1.0}
\definecolor{magenta}{rgb}{.1,0,.3}
\definecolor{gruen}{rgb}{0.2,0.5,.5}
\definecolor{light}{rgb}{ 0.992, 0.961,  0.902}
\definecolor{Tan}{rgb}{ 0.992, 0.9,  0.902}

\newcommand{\komment}[1]{{}}

\title{Contact Tracing \& Super-Spreaders in the Branching-Process Model}

\author{Johannes M\"uller         
	\and
	Volker H\"osel 
}

\institute{
Johannes M\"uller 
   \at Center for Mathematics, Technische Universit\"at M\"unchen, 85748 Garching, Germany and Institute for Computational Biology, Helmholtz Center Munich, 85764 Neuherberg, Germany
   \email{johannes.mueller@mytum.de}
\and 
Volker H\"osel 
   \at Center for Mathematics, Technische Universit\"at M\"unchen, 85748 Garching, Germany
} 
\date{Received: date / Accepted: date}

\begin{document}
\maketitle

\begin{abstract}
In recent years, it became clear that super-spreader events play an important role, particularly in the spread of  airborne  infections. We investigate a novel  model for super-spreader events, not based on a heterogeneous contact graph but on a random contact rate: Many individuals become infected synchronously in single contact events. We use the branching-process approach for contact tracing to analyze  the impact of super-spreader events on the effect of contact tracing. Here we neglect a tracing delay. Roughly speaking, we find that contact tracing is more efficient in the presence of super-spreaders if the fraction of symptomatics is small, the tracing probability is high, or the latency period is distinctively larger than the incubation period. In other cases, the effect of contact tracing can be decreased by super-spreaders. Numerical analysis with parameters suited for SARS-CoV-2 indicates that super-spreaders do not decrease the effect of contact tracing crucially in case of that infection.
\end{abstract}

\par\medskip 

\section{Introduction}

A visit to a restaurant, attending a church service, practicing a choir: The common feature of these occasions is that they all may lead to so-called super-spreader events. We start to understand the role of super-spreader events in the dynamics of an outbreak~\cite{Lloyd-Smith2005}. Depending on the transmission characteristics of a pathogen, the number of secondary cases per individual might be rather uniform, or disperse. In the disperse case, the majority of infecteds do not pass the infection at all, while some individuals become super-spreaders and infect many persons in their environment. That  heterogeneity might be caused by differences in the  immune response: Some individuals have a weak immune response and a high pathogen load. They are highly infectious. Another cause can be a heterogeneous contact network, where some individuals form highly connected hubs. These two aspects can be modeled by a static, heterogeneous contact graph. Another mechanism causing super-spreaders is based on a dynamic contact network. If a person is infectious but still in his/her incubation period and visits a restaurant, he/she might infect many individuals. If he/she stays at home during this decisive time interval, then only few or no further persons will be infected. The contact rate and contact structure is a major reason for the dispersion in the distribution of the individual reproduction number.
\par\medskip 
Super-spreaders attract attention in the recent literature. Data clearly indicate the existence for super-spreaders for several infections, as tuberculosis~\cite{Walker2013,Andre2007,Melsew2019}, SARS~\cite{Lloyd-Smith2005,AlTawfiq2020}, SARS-CoV-2~\cite{Liu2020a,AlTawfiq2020}, among others~\cite{Lloyd-Smith2005}. The scientific community develops models to describe super-spreaders. There are as simple models as deterministic dynamical systems of an SIR-type with different classes of infecteds~\cite{Mkhatshwa2010}. Several models are  based on the heterogeneity of contact graphs as small-world-networks~\cite{Small2006}, or aim to address social structure in an individual-based model~\cite{Duan2013}. The seminal work by Lloyed-Smith et al.\ did highlight the importance of super-spreading for infection dynamics and control~\cite{Lloyd-Smith2005}.\\

The present study aims to investigate the effect of super-spreading on contact tracing (CT). Often, CT is modeled by individual or agent based based models (e.g.~\cite{Tian2011,Liu2015,kiss2008}). These models allow to include many effects, e.g.\ a detailed social contact graph, or the implementation of sophisticated tracing protocols. The draw-back is the fact that they are investigated by simulations only, such that the dependencies of the outcome on parameters are not clearly visible. It is also difficult to obtain more general rules from this approach. Often, these models are used to investigate a given infection in a given situation. A second approach is based on pair approximation (e.g.~\cite{Keeling1999,eames2002,Huerta2002,house2010}) -- a stochastic, individual based model is approximated by a system of ordinary differential equations (ODE's). 
The advantage of this method is that it is based on first principles, but the number of ODE's necessary in pair approximation is roughly the squared number of the states an individual can assume, and in this, the resulting ODE's are high dimensional and hard to analyses analytically. Numerous simple ODE models for CT aim at a more handy structure~\cite{Arazoza2002,Hsieh2010,Heffernan2009}. However, as these models are based on {\it ad hoc} assumptions and not at first principles, it is in general  rather difficult to relate the results to CT on a micro scale. The present study follows the branching-process approach proposed by~\cite{mueller2000,ball2011}. The tree of infecteds is investigated: The nodes of this tree are the infected individuals, an directed edge goes from infector to infectee. If an individual is diagnosed, it becomes an index case. The neighboring nodes have probability $p$ to be traced. All persons who are detected go in quarantine and are assumed not to be infectious any more. This approach allows for an analytical treatment, though the resulting mathematical structure is not as handy as the ODE models mentioned above. Therefore, Browne at al.~\cite{Browne2015} take up this approach and formulate an approximate ODE model based on these ideas.\\

Close to the analysis of super-spreader events on the effect of CT is the analysis of the interplay between the heterogeneity in the contact graph and CT~\cite{eames2007contact,house2010}. Particularly the nice simulation study by Kiss et al.~\cite{kiss2008} focuses on assortatively and disassortatively mixing contact graphs,  which resembles the consideration of dispersion in the contact structure. In the same spirit, Okolie et al.~\cite{Okolie2020} formulate the branching  process approach on a random tree, where also the influence of the variance in the degree structure is discussed. Eames and Keeling~\cite{eames2003contact} find a formula for the critical tracing probability, and note that this formula is valid under many circumstances, also for super-spreading. Hyman et al.~\cite{Hyman2003} use a deterministic model to conclude that CT is more efficient to find super-spreaders than screening. This finding is confirmed in~\cite{Kojaku2020}, where a branching process on a network is considered analytically; the authors show the stronger conclusion that CT even is superior to acquaintance sampling. However, the two paper do not compare the effect of CT in presence and absence of supers-spreading events. Klinkenberg at al.~\cite{klinkenberg2006} argue that large infection events can be readily detected, and in this, CT is effective also in this case. The paper~\cite{Reich2020} formulates a small-world-network allowing for super-spreader events for SARS-CoV-2, and assesses control strategies using the model.   All in all, most of the studies about CT and super-spreaders published so far prescribe a contact graph, and in that, assume implicitly that the properties of given individuals (as the heterogeneity in the immune system or the heterogeneity in the contact structure) cause  super-spreader events.\par\medskip 
In contrast, the present paper rather focuses on the hypothesis that any person may become a super-spreader, just by chance (the famous visit of a birthday party or of a restaurant). The main difference to the heterogeneous-network-approach is that many individuals are infected at the very same time. Since CT is based on subtle timing (a race between infection and detection), we expect that this aspect might be of importance. In our model, the population and the contact structure is homogeneous, but the contact rate is assumed to be random. We extend the branching process approach for CT to cover contact rates that are random functions of a certain  class. The theory developed is used to investigate quantitatively the effect of dispersion/super-spreader events on the efficiency of CT. We find that CT might be more or less efficient for super-spreader events, depending on the parameters. Moreover, the mechanism that allows CT to control an infection is different for the two cases: without super-spreader events, it rather is based on preventing further infection, while in super-spreader events it is more based on the diagnose of part of the persons infected in the event. For SARS-CoV-2 we can show that the influence of super-spreader events on the efficiency of CT only is gradual and will not be decisive in the fight against the infection.

\section{Model}
We focus on the onset of an outbreak in a large, homogeneous population and consider an S(E)I$\ast$ model, where $\ast$ represents any non-infectious state, as R or S. As usual in this setting, it is appropriate to formulate the model as a branching process~\cite{ball:donnelly}. Direct interaction between infected individuals or infected and recovered individuals can be neglected. \\
An infected person recovers spontaneously (without diagnosis) at rate $\mu(a)$, and develops symptoms and becomes diagnosed at rate $\sigma(a)$, where $a$ denotes the age (time) since infection (a.s.i.). An infectious person has  contacts at rate $\beta(a)$. Usually, $\beta(a)$ is a deterministic function, that is, the number of infectious contacts in the interval $[0,a]$ follows a Poisson distribution with expectation $\int_0^a\beta(\tau)\, d\tau$. In order to model super-spreader events, we divide the contact rate in a deterministic part (``deterministic contacts'') $\beta_0(a)$ and a random part. The deterministic part works as usual. For the random part (``random contacts''), we define a Poisson process with arrival times $(T_i)_{i\in\N}$ and corresponding counting process $Y_a=\#\{T_i<a\}$. This Poisson process is homogeneous with rate $\lambda$. Super-spreader events may take place on these time points $T_i$: The number of secondary cases produced at those discrete time points follows a Poisson distribution with expectation $\beta_1(T_i)/\lambda$. As we will find out, the scaling of $\beta_1(a)$ by $1/\lambda$ yields a reproduction number which is independent of $\lambda$. Let $Z(u), Z_a(u)\sim \mbox{Pois}(u)$, where $Z(u)$, $Z_a(u)$ and $Z_{a'}(u)$ are independent (for $a\not= a'$). 
If an individual is infectious in the a.s.i-interval $[0,a_0]$, the number of secondary cases produced in that interval is given by 
\begin{eqnarray}\label{xContactProcess}
 Z\bigg(\int_0^{a_0}\beta_0(a)\, da\bigg) + \sum_{i=1}^\infty Z_{T_i}(\beta_1(T_i)/\lambda)\eins(T_i<a_0)
= Z\bigg(\int_0^{a_0}\beta_0(a)\, da\bigg) + \int_0^{a_0} Z_a(\beta_1(a)/\lambda) dY_a
.\quad 
\end{eqnarray}
We assume $\mu(a),\sigma(a),\beta_i(a)\in C^0(\R_+,\R_+)$, and that there are $\varepsilon, \overline a>0$ s.t. $\mu(a)+\sigma(a)\geq\varepsilon$ for $a>\overline a$. For convenience, we introduce the random function $\beta(a)$ 
\begin{eqnarray}
 \beta(a) = \beta_0(a) + \sum_{i=1}^\infty Z_{T_i}(\beta_1(T_i)/\lambda)\,\delta_{T_i}(a),
\end{eqnarray}
with the understanding that $\beta(a)$ indicates the number of contacts for a given individual as given by  eqn~\ref{xContactProcess}. 
Here, $\delta_T(a)$ refers to a Dirac delta (point mass) at $a=T$. As $\beta(a)$ is the direct generalization of the infection rate to the situation at hand, we will also call $\beta(a)$ the ``infection rate''.

\par\medskip 
\begin{figure}[htbp]
\begin{center}
	\includegraphics[width=5cm]{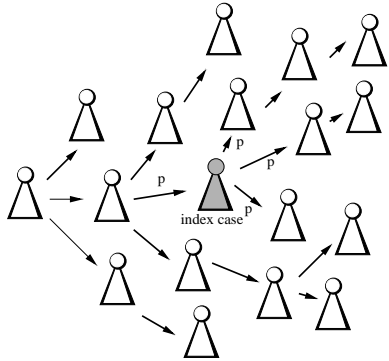}
\end{center}	
\caption{Sketch of the concept for CT.}
\label{ctSketch}
\end{figure}
The (random) contact rate $\beta(a)$ and the removal rate $\mu(a)+\sigma(a)$ define a branching process~\cite{Athreya1972}. 
This branching process, in turn, defines the tree of infecteds: The nodes are infected individuals, a directed edge goes from infector to infectee. Note that each infector has its own realization of the contact process $\beta(a)$, independently of all other individuals in the tree of infecteds. The tree rather is a forest, as recovered individuals leave this tree. Particularly, we assume that recovered individuals never form index cases for CT. Index cases are individuals who show symptoms and are diagnosed. The infectious contacts, that is, the neighbors of the index case in the tree of infecteds, have probability $p$ to be traced (see Fig.~\ref{ctSketch}). In the present paper, we do not take a tracing delay into account (find in~\cite{muller2016effect} results for the tracing delay without random contacts). In one-step tracing, only the direct neighbors can be traced. In recursive tracing, the detected persons also become index cases, and snowballing is triggered. We call this family of processes ``branching-tracing processes''.

\subsection{Distribution of $R_{ind}$}
A central aspect of the present study is the investigation of the dispersion factor and its effect on CT. Therefore, we first investigate the distribution of $R_{ind}$ without CT, where $R_{ind}\in\N_0$ denotes the random variable of the number of secondary infecteds for the different  individuals. Accordingly, the reproduction number is defined by $R_0=E(R_{ind})$. The dispersion factor describes the heterogeneity of $R_{ind}$. If all individuals behave similar, we expect the distribution of the number of secondary cases to be geometric (constant parameters, exponentially distributed infectious period) or Poissonian (constant infection rate, infectious period is fixed). While the expectation is $R_0$, the variance is $R_0(1+R_0)$ (for the geometric distribution) or $R_0$ (for the Poisson distribution). In case of super-spreader events, the variance is increased.  Following~\cite{Lloyd-Smith2005}, we take a negative Binomial distribution as background, and define the dispersion factor $k$ by the relation between variance and expectation of $R_{ind}$, 
\begin{eqnarray}\label{xxKequEins}
 \mbox{var}(R_{ind}) = R_0 (1+R_0/k).
\end{eqnarray}
Hence, $k=\infty$ for the Poisson distribution, and $k=1$ for the Geometric distribution. If the variance is even higher, $k$ becomes smaller. That is, the smaller $k$, the larger the variance, and the higher the variability among the number of secondary cases produced by an individual. However, it is to mention that eqn.~(\ref{xxKequEins}) simply is a  definition for $k$, and that there are different definitions to characterize dispersion. 
\begin{ddef} The dispersal factor $k$ for $R_{ind}$ is defined by 
\begin{eqnarray}
k = \frac{R_0^2}{\mbox{var}(R_{ind})-R_0}.
\end{eqnarray}
\end{ddef}
Note that for our model, we always have $\mbox{var}(R_{ind}) > R_0$. In case of constant parameters, this inequality is a  consequence of proposition~\ref{varProp}. 

\begin{theo} Let $A$ be an $\R_+$-valued random variable with
$ P(A>a) = e^{-\int_0^a\mu(\tau)+\sigma(\tau)\, d\tau}$.  With the notation introduced above,
$$R_{ind}\, = \,Z\bigg(\int_0^A\beta_0(a)\, da\bigg) + \int_0^A Z_a(\beta_1(a)/\lambda)\, dY_a$$
and 
$$R_0 = E(R_{ind}) = \int_0^\infty (\beta_0(a)+\beta_1(a))\,e^{-\int_0^a\mu(\tau)+\sigma(\tau)\, d\tau}\, da. 
$$
\end{theo}
{\bf Proof: }
If we consider a given individual, his/her infectious period is distributed according to the random variable $A$. 
The formula for $R_{ind}$ is a direct consequence of the way how we did construct the contact process. We take the expectation of $R_{ind}$. According to the law of iterated expectations, we have $R_0=E(E(R_{ind}|A))$, and therewith we obtain
\begin{eqnarray*}
R_0 &=& 
\int_0^\infty E\bigg(Z\bigg(\int_0^a\beta_0(\tau)\,d\tau\bigg)\,
+
\int_0^a Z_\tau(\beta_1(\tau)/\lambda)\, dY_\tau
\bigg) (\mu(a)+\sigma(a))\,e^{-\int_0^a\mu(\tau)+\sigma(\tau)\, d\tau}\, da  \\
& = &  
 -\int_0^\infty  \bigg[\int_0^a\beta_0(\tau)\, d\tau + \frac 1 \lambda E\bigg(\int_0^a \beta_1(\tau)\, dY_\tau\,\bigg)\bigg] \frac d{da}e^{-\int_0^a\mu(\tau)+\sigma(\tau)\, d\tau}\,da\\
& = & \int_0^\infty \beta_0(a)\, e^{-\int_0^a\mu(\tau)+\sigma(\tau)\, d\tau}\, da  
+\frac 1 \lambda E\bigg(\int_0^\infty  \beta_1(a)\,  e^{-\int_0^a\mu(\tau)+\sigma(\tau)\, d\tau}\,dY_a\bigg). 
\end{eqnarray*}
We 
recall that the probability density  of $T_i$ is given by $\frac{\lambda}{(i-1)!}(\lambda t)^{i-1}e^{-\lambda t}$ for $t\geq 0$, 
and proceed
\begin{eqnarray*}
	&&\frac 1 \lambda E\bigg(\int_0^\infty  \beta_1(a)\,  e^{-\int_0^a\mu(\tau)+\sigma(\tau)\, d\tau}\,dY_a\bigg) 
 =  
\frac 1 \lambda \sum_{i=1}^\infty E\bigg(  \beta_1(T_i)\,  e^{-\int_0^{T_i}\mu(\tau)+\sigma(\tau)\, d\tau}\,\bigg)\\
 &= & \frac 1 \lambda \sum_{i=1}^\infty \int_0^\infty 
 \,\beta_1(a)\,  e^{-\int_0^{a}\mu(\tau)+\sigma(\tau)\, d\tau}\, \frac{\lambda\,(\lambda a)^{i-1}}{(i-1)!}e^{-\lambda a}da 
 =  \int_0^\infty 
\,\beta_1(a)\,  e^{-\int_0^a\,\mu(\tau)+\sigma(\tau)\, d\tau}\,da.
\end{eqnarray*}
\qed\par\medskip

Next we aim at an explicit expression for the dispersion coefficient in case that the parameter functions $\mu$, $\sigma$, $\beta_0$, and $\beta_1$ are independent of $a$ (are constant), s.t.\ 
$ R_{ind} =
	Z(\beta_0 A)\,+\,\int_0^A Z_a(\beta_1/\lambda)dY_a.$ \\
	
\begin{prop} \label{varProp} In case of constant parameter functions, we have 
$$\mbox{var}(R_{ind}) = R_0(1+R_0) + \frac{\beta_1^2}{\lambda\,(\mu+\sigma)}.$$
\end{prop}
{\bf Proof: }
In order to obtain the variance of $R_{ind}$ we use Eve's law: For two random variables $X$ and $Y$ we have  
$\mbox{var}(X) =  \mbox{var}(E(X|Y)) + E(\mbox{var}(X|Y))$. 
We start with (let $a\in\R_+$ be given, fixed) 
\begin{eqnarray*}
\mbox{var}\bigg(\sum_{i=1}^{Y_a}Z_i(\beta_1/\lambda)\bigg) &=&
\mbox{var}\bigg(E\bigg(\sum_{i=1}^{Y_a}Z_i(\beta_1/\lambda)\bigg|Y_a\bigg)\bigg) 
+
E\bigg(\mbox{var}\bigg(\sum_{i=1}^{Y_a}Z_i(\beta_1/\lambda)\bigg|Y_a\bigg)\bigg)\\
&=& \mbox{var}(Y_a)(\beta_1/\lambda)^2 + E(Y_a)(\beta_1/\lambda) =  \beta_1^2\,a/\lambda +  a\beta_1.
\end{eqnarray*}
Therewith,
\begin{eqnarray*}
E(\mbox{var}(R_{ind}|A)) 
&=& E\bigg(\mbox{var}\bigg(Z(\beta_0 A)\,+\,\int_0^A Z_a(\beta_1/\lambda)dY_a\bigg|A\bigg)\bigg) 
= 
E\bigg(\beta_0\,A + \mbox{var}\bigg(\sum_{i=1}^{Y_A}Z_i(\beta_1/\lambda)\bigg| A\bigg) \bigg)\\
& = &  E\bigg(A(\beta_0+\beta_1) + \beta_1^2\,A/\lambda\bigg) 
= R_0 + \frac{\beta_1^2}{\lambda\,(\mu+\sigma)}.
\end{eqnarray*}
The last ingredient is the computation of $\mbox{var}(E(R_{ind}|A))$,
\begin{eqnarray*}
\mbox{var}(E(R_{ind}|A)) &=& \mbox{var}((\beta_0+ \beta_1) A) =  R_0^2.
\end{eqnarray*}
Hence Eve's law yields the desired result.
\qed\par\medskip 

In the next corollary, we state the dispersion factor for the special case of our model (recall that $R_0 = (\beta_0+\beta_1)/(\mu+\sigma)$).
\begin{cor}
If the rates are constant, the dispersion factor for $R_{ind}$ is given by
\begin{eqnarray}
 k = \left(1+\frac{\beta_1^2}{(\beta_0+\beta_1)^2}\,\,\frac{\mu+\sigma}{\lambda} \right)^{-1}. 
\end{eqnarray}
\end{cor}

Particularly if $\lambda$ becomes large, the events of the 
Poisson process become frequent. If the time between two subsequent superspreader events in the contact rate typically is much shorter than the infectious period ($\mu+\sigma\ll\lambda$), the discrete events are practically averaged out (as in a moving average), and the model approximates a situation with an effective deterministic contact rate $\beta_0+\beta_1$, and $k$ tends to $1$. Only if $\lambda$ is distinctively smaller than $\mu+\sigma$, single (and w.r.t.\ the time scale of the infectious period) seldom super-spreader events take place. In this case, $k$ becomes small.
\par\medskip 
We proceed to investigate the distribution of $R_{ind}$ in the case of constant parameters.

\begin{theo}  If all parameters are independent of $a$, the generating function of $R_{ind}$ reads
	$$\varphi_{R_{ind}}(s) = \frac{\mu+\sigma}{\mu+\sigma + \left(1-e^{(s-1)\beta_1/\lambda}\right)\lambda+(1-s)\beta_0}.$$
\end{theo}
{\bf Proof: }
Let again $A\sim\mbox{Exp}(\mu+\sigma)$ denote the infectious period of a focal individual. If we condition 
on $A=a$, then the number of concactees from the deterministic part  (rate $\beta_0$) follows a Poisson distribution with expectation $\beta_0 a$ and generating function $\varphi_0(s;a) = e^{(s-1)\beta_0 a}$. The number of contactees due to the random part can be written as
$$ \sum_{i=1}^{Y_a} Z_i$$
where $Y_a$ is the counting process for the Poisson process, and $Z_i\sim\mbox{Pois}(\beta_1/\lambda)$. 
That is, we have a compound random variable. The generating function is a concatenation of the generating function for $Y_a$ and for $Z_i$, s.t. the generating functions of that part reads
$$ \varphi_1(s;a) = e^{(e^{(s-1)\beta_1/\lambda}-1)\lambda a}.$$
As the two processes to generate infectees are independent if we condition on $A$, the generating function of the sum is the product of $\varphi_1(s)$ and $\varphi_2(s)$. All in all, we obtain 
$$ \varphi_{R_{ind}}(s) = \int_0^\infty \varphi_0(s;a)\, \varphi_1(s;a)\, (\mu+\sigma)e^{-(\mu+\sigma)a}\, da =  \frac{\mu+\sigma}{\mu+\sigma + \left(1-e^{(s-1)\beta_1/\lambda}\right)\lambda+(1-s)\beta_0}.$$
\qed\par\medskip 

\begin{rem}\label{majorOurbreak}
We find that 
$$ \lim_{\lambda\rightarrow 0} \varphi_{R_{ind}}(s) =  \frac{\mu+\sigma}{\mu+\sigma +(1-s)\beta_0}$$
and 
$$ \lim_{\lambda\rightarrow \infty} \varphi_{R_{ind}}(s) =  \frac{\mu+\sigma}{\mu+\sigma +(1-s)(\beta_0+\beta_1)}$$
where the convergence is pointwise. 
The probability for a major outbreak (if we start with one infected individual) does only weakly depend on the random contact events if $\lambda$ is small,  while for frequent random events, the probability for extinction is given by the fixed point of the generating function (we have a Galton-Watson-Process), that is,  by $1/R_0$ (in case of $R_0>1$).\\
For $\lambda\rightarrow\infty$, the model  approximates a situation without superspreading and a total contact rate $\beta_0(a)+\beta_1(a)$. That is in line with our expectations, as in this case the random ``superspreader events'' are tiny but frequent, and in that, resembles the way usually contacts are modeled.\\ 
Particularly interesting is the fact that the random events do contribute to $R_0$ (and also influence the exponential growth rate of the epidemic in the onset), but in the limit $\lambda\rightarrow 0$, the probability for extinction $q^*$ is given by $(\mu+\sigma)/\beta_0$ (if this expression is less than $1$), such that $\beta_1$ does not play a role. 
We can understand this finding intuitively. If $\lambda$ is close to zero, the random contact events are that seldom, that in the initial time interval of the outbreak (which is decisive for the dichotomy: to die out or to generate a major outbreak) it is very, very unlikely that a random contact event takes place. Therefore, these events practically do not play a role in the probability of a major outbreak. Mainly the deterministic part of the contact structure determines this probability, though the random and the deterministic part of the contact structure contribute equally to the reproduction number. That is an important finding, as we will rediscover this phenomenon below in the investigation of backward tracing. \\
\end{rem} 

\section{CT -- Model analysis}

In the present section we aim to analyze the tracing-branching process. The approach is based on the computation of the marginal probability that a given individual is infectious at a certain age of infection. In a computational approach, we would run the stochastic process very often, select a certain focal individual (e.g. the primary infected person), and determine the fraction of realizations in which this individual is still infectious at age $a$ of infection. This probability is a marginal probability -- we average over the state of all other individuals. In that, we remove most of the  correlations that make the analysis difficult or even  intractable. As we will see, we are nevertheless able to work out the expected effective reproduction number of our focal individual (how many individuals in average our focal individual did infect). However, as different generations are not independent, as it is the case for a standard branching process, the usual argument via the embedded Galton-Watson process to obtain a threshold theorem (reproduction number smaller $1$ will lead to the extinction of the process a.s., if the reproduction number is larger one, we have persistence of the process with a positive probability) is not feasible. Therefore, we add at the end of the present section a discussion the relevance of our findings.

\subsection{Preliminaries}
Before we start with the analysis of the tracing model, we aim to clarify how to compute expectations in certain transition models with rate constants that are random variables in themselves.\\ 
Let $Y_1$, $Y_2$ be independent, exponentially distributed random variables, $Y_i\sim\mbox{Exp}(\mu_i)$ for $i=1,2$. Consider a particle that is in state A at time $t=0$, and jumps to state $B$ depending on $Y_1$ and/or $Y_2$ (see below). Let furthermore $X_t=1$ if the particle is in state A, and $X_t=0$ if it is in state B (at time $t$). \\
If the particle jumps at time $Y_1$, then 
$$ \frac d{dt} P(X_t=1) = -\mu_1 P(X_t=1).$$
If the particle jumps at time $Y_2$, we find similarly
$$ \frac d{dt} P(X_t=1) = -\mu_2 P(X_t=1).$$
If the particle jumps at $Y_1$ or $Y_2$ (which time is earlier), then 
$$ \frac d{dt} P(X_t=1) = -(\mu_1 + \mu_2) P(X_t=1).$$
Now we introduce a random variable $\eta$,  independent of $Y_i$, and $P(\eta=1)=P(\eta=2)=1/2$. If the particle jumps at $Y_\eta$ (we select $Y_1$ and $Y_2$ at equal probability), what is the right equation? Clearly, 
$$ P(X_t=1) = \frac 1 2 e^{-\mu_1 t}+\frac 1 2 e^{-\mu_2 t}.$$
If we consider $\mu=\mu_\eta$ as a random rate, we may write
$$ P(X_t=1) = E_\mu\bigg( e^{-\mu t}\bigg).$$
Now we change the game slightly. In each time interval $\Delta t$, we start with probability $\beta \Delta t$ a clock. All time intervals are handled independently, and all clocks follow (independently) an exponential distribution with rate $\mu$. A clock that rings will trigger the transition of the particle with probability $p$. We can reformulate the model at hand as an immigration-death process (or an M/M/$\infty$ queue), where immigrants arrive independently at rate $\beta$, die at rate $\mu$, and a death event is observed with probability $p$. $X_t$ is $1$ until the first observed death event of an immigrant happens.\\
Assume that no clock did ring before $t$. The probability that a clock was started in $[c,c+\Delta t]$ and rings in $[t,t+\Delta t]$ is then given by 
$$ \beta\Delta t\,\,e^{-\mu\, (t-c)}\,\mu\Delta t.$$
In order to obtain $P(X_{t+\Delta t}=1)$ in terms of the history of $X_t$, we consider the process on a grid with step width $\Delta t$ (below we will sum over all grid points $c$  with $c\leq t$), and take the limit $\Delta t\rightarrow 0$ afterwards. We obtain
\begin{eqnarray*}
P(X_{t+\Delta t}=1) 
&=& P(X_t=1) - P(\mbox{ a clock rings and is successful }|X_t=1)\,P(X_t=1)+{\cal O}(\Delta t^2)\\
&=& P(X_t=1) -\bigg( \sum_{c\leq t} p\,\beta\Delta t\,\,e^{-\mu\, (t-c)}\,\mu\Delta t\,\bigg)\,P(X_t=1)+{\cal O}(\Delta t^2)\\
\Rightarrow\quad  \frac d {dt}P(X_t=1) 
 &=& -\int_0^t p\beta e^{-\mu\, (t-c)}\,\mu\, dc\,P(X_t=1)
 = -\int_0^t p\beta \frac d {dc}e^{-\mu\, (t-c)}\, dc\,P(X_t=1)
\end{eqnarray*}
It is obviously possible to simplify the integral. Instead, we use two types of clocks (type $Y_1$ and type $Y_2$) and decide (independently)  at each time point if we start a clock (probability $\beta\Delta t$) and -- if we start a clock -- which type of the clock we want to use.  
If both types have the same chance to be used, we have 
(recall that $\eta=\eta_\eta$ is a random variable)
\begin{eqnarray*}
	\frac d {dt}P(X_t=1) 
	&=& -\int_0^t p\beta \frac d {dc} E_\mu(e^{-\mu\, (t-c)})\, dc\,\,\,P(X_t=1)\\
\Rightarrow\quad 	P(X_t=1)  &=& 
	\exp\bigg\{ -\int_0^t\int_0^s p\beta \frac d {dc} E_\mu(e^{-\mu\, (s-c)})\, dc\, ds\bigg\}
\end{eqnarray*}
Last, we also convert $\beta$ into a random variable. Then, 
\begin{eqnarray}\label{expectProto}
	P(X_t=1)  &=& 
E_\beta\bigg(\exp\bigg\{ -\int_0^t\int_0^s p\beta \frac d {dc} E_\mu(e^{-\mu\, (s-c)})\, dc\, ds\bigg\}\bigg).
\end{eqnarray}
For our analysis below it is important to note that the expectation w.r.t.\ $\beta$ is outside of the exponential function, while the expectation w.r.t. $\mu$ is inside the exponential function. 

\subsection{Backward tracing}
\begin{figure}[htbp]
	\begin{center}
		\includegraphics[width=5cm]{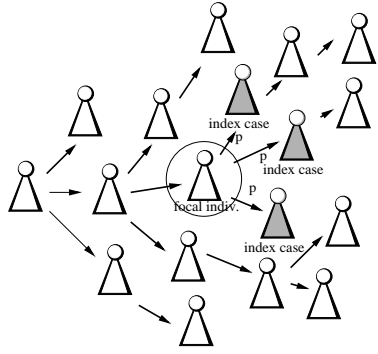}
	\end{center}	
	\caption{Sketch of the concept for backward tracing. Only if an infectee (in gray) becomes an index cases, the focal individual can be traced.}\label{ctSketchBack}
\end{figure}

\begin{figure}[htbp]
	\begin{center}
		\includegraphics[width=5cm]{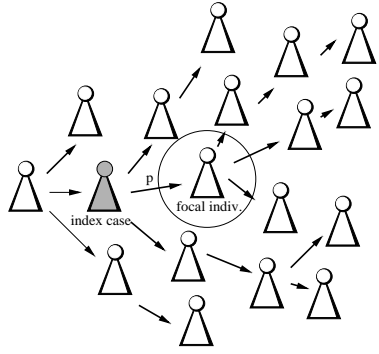}
	\end{center}	
	\caption{Sketch of the concept for forward tracing. Only if the infector becomes an index cases, the focal individual can be traced.}\label{ctSketchForw}
\end{figure}

Following the papers~\cite{mueller2000,muller2016effect,mueller2007,Okolie2020}, we first distinguish between backward tracing (a person can be only traced via his/her infectees, Fig~\ref{ctSketchBack}), forward tracing (a person is only traced by his/her infector, Fig~\ref{ctSketchForw}), and full tracing, where tracing via infector and infectee is possible (Fig~\ref{ctSketch}). We start with backward tracing.\par\medskip 

We allow for age-dependent rates $\beta_0(a)$, $\beta_1(a)$,  $\mu(a)$ and $\sigma(a)$ and define 
$$  \widehat\kappa(a) = \exp\bigg(-\int_0^a\mu(\tau)+\sigma(\tau)\,d\tau\bigg)
.$$
In the analysis, we need to be clear about dependencies, particularly dependencies w.r.t.\ the contact process. In Fig.~\ref{ctSketchBack}, each individual has his/her own i.i.d.\ copy of the contact process. The number of infectees depend on the contact rate of an infector. As we consider backward tracing, also a focal individual's probability to be infectious at a given age-of-infection is influenced by its own contact process. For the realized infectees, however, the probability to be infectious at a.s.i.\ $a$ does not depend on the infector's contact process.\\
In order to better express this fact in the analysis, we distinguish in the proof below between $\kappa^-_0(a)$, that is the probability of a focal individual to be infectious at a.s.i.\  $a$, {\it given the realizations of its own contact process}, and $\ol\kappa^-_0(a)$, where we {\it average over all possible realizations of the contact process}, $\ol\kappa^-_0(a)=E(\kappa^-_0(a))$.

\begin{prop}\label{backExpectRecProp} 
 For recursive tracing, we find  
\begin{eqnarray}\label{backExpectRec}
\overline \kappa_0^-(a) = \widehat\kappa(a)\, 
E\bigg(\exp\bigg\{-p\,\int_0^a\int_0^s\beta(a-c)\,(-\overline\kappa_0^-{'}(c)-\mu(c)\overline\kappa_0^-(c))\,dc\,ds\bigg\}\bigg).
\end{eqnarray}
\end{prop}
{\bf Proof: } We first look at one given realization of all contact rates.  Only afterwards, we take the appropriate expectations. 
Without CT, an individual is infectious at a.s.i.\  $a$ with probability $\widehat\kappa(a)$. 
This probability is decreased by CT. 
Only infectees cause tracing, hence the probability to be infectious at a given a.s.i.\  $a$ has the same expectation for infector and infectee. As the realization of the contact rate for each infectee may be different, we add a tilde to their survival probabilities. Note that each infectee with different a.s.i.\  is another individual, with his/her own (independent) set of contact realizations. In backward tracing, the location of an individual in the tree of infecteds does not matter. In this sense, the recovery rate of an infectee is given by the hazard rate 
$$ \frac{-{{\tilde \kappa}^-_0}{'}(b)}    
       {{\tilde \kappa}^-_0(b)}.$$
This hazard rate includes the rate of spontaneous recovery $\mu(a)$, the rate of direct observation $\sigma(a)$, but also the removal rate due to CT.
 The rate to be detected (indirectly or directly) is given by 
$$ \frac{-{{\tilde \kappa}^-_0}{'}(b)}
{{\tilde \kappa}^-_0(b)}-\mu(b).$$
The focal individual (for which we compute $\kappa^-(a)$) produces during his/her complete infectious time span $[0,a]$ infectees. When he/she has had age $c\in[0,a]$, the expected number of secondary cases per time interval was $\beta(c)$ (here we condition on the infector's Poisson process $(T_i)_{i\in\N}$). 
The probability that the infectee is now (infector has age $a$) still infectious is $\tilde \kappa^-_0(a-c)$. The rate of direct or indirect observation of an infectee with a.s.i.\  $a-c$  is 
$ \frac{-{\tilde \kappa^-_0}{'}(a-c)}{\tilde \kappa^-_0(a-c)}-\mu(a-c)$.
A detected individual triggers a successful tracing event with probability $p$. Hence, the contribution of the removal rate of our focal individual due to tracing rate is given by
$$ p\,\int_0^a \beta(c)\,
\tilde\kappa^-_0(a-c)\bigg(\frac{-{\tilde\kappa^-_0}{'}(a-c)}{\tilde\kappa^-_0(a-c)}-\mu(a-c)\bigg)
\, dc.$$
We find for the given realization of the contact rates
$$
\frac d {da}\kappa_0^-(a) = 
- \kappa_0^-(a)\bigg\{
\mu(a)+\sigma(a) + p\,\int_0^a \beta(c)\,
\tilde\kappa^-(a-c)\bigg(\frac{-{\tilde\kappa^-_0}{'}(a-c)}{\tilde\kappa^-_0(a-c)}-\mu(a-c) \bigg)\,dc\,
\bigg\},\quad \kappa^-_0(0)=1.
$$
Therefrom we obtain 
$$
 \kappa_0^-(a) = \widehat\kappa(a)\, 
\exp\bigg\{-p\,\int_0^a\int_0^s\beta(a-c)\,(-\tilde\kappa_0^-{'}(c)-\mu(c)\tilde\kappa_0^-(c))\,dc\,ds\bigg\}
$$
Note that $\beta$ is independent of $\tilde\kappa_0^-(a)$, as 
$\beta$ is the contact rate of the focal individual, and $\tilde\kappa_0^-$ is the survival probability for a given infectee of the focal individual. According to the considerations above, particularly eqn.~(\ref{expectProto}), the expectation w.r.t. (all) contact rates involved can be written as 
$$\overline \kappa_0^-(a) = \widehat\kappa(a)\, 
E\bigg(\exp\bigg\{-p\,\int_0^a\int_0^s\beta(a-c)\,(-\overline\kappa_0^-{'}(c)-\mu(c)\overline\kappa_0^-(c))\,dc\,ds\bigg\}\bigg).$$
\qed\par\medskip

\begin{prop} For one-step tracing, we find  
	\begin{eqnarray}\label{backExpectOneStep}
\overline	\kappa_0^-(a) = \widehat\kappa(a)\, 
E\bigg(\exp\bigg\{-p\,\int_0^a\int_0^s\beta(a-c)\,
      \sigma(c)\overline \kappa_0^-(c)\,dc\,ds\bigg\}\bigg).
	\end{eqnarray}
\end{prop}
{\bf Proof: } The proof parallels that of proposition~\ref{backExpectRecProp}, only that we replace the expression 
$$ \frac{-\kappa_0^-{'}(c)}{\kappa_0^-(c)}-\mu(c)$$
by the rate of direct observations $\sigma(c)$.
\qed \par\medskip 

The next aim is to remove the expectation from the formulas (\ref{backExpectRec}) and (\ref{backExpectOneStep}). 
Let $W_\lambda(z)$ denote the probability generating function for the Poisson distribution, 
$$W_\lambda (z) = e^{\lambda(z-1)}
$$
In the proof of the next theorem we use a well known, handy lemma about the conditioned Poisson process~\cite[page 612]{Shanbhag2001}.
\begin{lem}\label{lemmaPoissonRandom}
	Consider the Poisson process $(T_i)_{i\in\N}$ with counting process $Y_a$, and for $k\in\N_0$
	$$\Omega_{a,k} = \{T_1,\ldots,T_k\,|\,Y_a=k\}.$$ 
	Let furthermore  $\tilde \Omega_{a,k}$, consisting of $k$ i.i.d. copies of random variables uniformly distributed in $[0,a]$. 
	Then,  
	$\Omega_{a,k}$ and $\tilde \Omega_{a,k}$ have 
	the same distribution, and for $g\in C^0(\R)$
	$$ E\bigg(\sum_{T\in\Omega_{k,a}} g(T)\bigg) 
	= E\bigg(\sum_{T\in\tilde \Omega_{k,a}}g(T)\bigg) = \frac k a\,\int_0^a g(t)\, dt.$$
\end{lem}
We are now equipped to prove the following theorem.

\begin{theo}\label{backwardTheorem} Given the infection rate, i.i.d. for each infected individual, 
$$  \beta(a) = \beta_0(a) + \sum_{i=1}^\infty Z_{T_i}(\beta_1(T_i)/\lambda)\,\delta_{T_i}(a),$$
where $(T_i)_{i\in\N}$ is a Poisson process with rate $\lambda$, 
we obtain for recursive tracing
\begin{eqnarray}\label{kappaMinusRekurs}
\overline	\kappa_0^-(a) 
&=& \widehat\kappa(a)\, 
e^{ -p\int_0^a\,(a-c)\,\beta_0(a-c)\,
	(-\overline\kappa_0^-{'}(c)-\mu(c)\overline\kappa_0^-(c))\,dc}\,\, \\
&&
\times  W_{\lambda\,a}\bigg(\frac 1 a \int_0^a\exp\bigg\{\bigg[e^{-p\,
(a-c)\,
(-\overline\kappa_0^-{'}(c)-\mu(c)\overline\kappa_0^-(c))}
-1\bigg]
\,\beta_1(a-c)\lambda^{-1}
\,\bigg\}\, dc\bigg).\nonumber 
\end{eqnarray}
and for one-step tracing
\begin{eqnarray}\label{kappaMinusOneStep}
\overline	\kappa_0^-(a) 
&=& \widehat\kappa(a)\, 
e^{ -p\int_0^a\,(a-c)\,\beta_0(a-c)\,
	\sigma(c)\overline \kappa_0^-(c)\,\,dc}\,\, \\
&&
\times  W_{\lambda\,a}\bigg(\frac 1 a \int_0^a\exp\bigg\{\bigg[e^{-p(a-c)\sigma(c)\ol\kappa_0^-(c)}-1\bigg]\beta_1(a-c)/\lambda\,\bigg\}\, dc\bigg).\nonumber 
\end{eqnarray}
\end{theo}
{\bf Proof: } We focus on one-step tracing, as the argument for recursive tracing is completely parallel. 
We may rewrite equ.~(\ref{backExpectOneStep}) as 
\begin{eqnarray*}
\overline	\kappa_0^-(a) 
&=& \widehat\kappa(a)\, 
E_{\beta}\bigg(\exp\bigg\{-p\,\int_0^a\,(a-c)\,\beta(a-c)\,
\sigma(c)\overline \kappa_0^-(c)\,dc\bigg\}\bigg). 
\end{eqnarray*}
Taking the given form of the contact rate into account, we find
\begin{eqnarray*}
&&\int_0^a\,(a-c)\,\beta(a-c)\,
\sigma(c)\overline \kappa_0^-(c)\,dc
= 
\int_0^a\,c\,\beta(c)\,\,
\sigma(a-c)\,\,\overline \kappa_0^-(a-c)\,dc\\
&=& \int_0^a\,(a-c)\,\beta_0(a-c)\,\,
\sigma(c)\,\,\overline \kappa_0^-(c)\,dc
+
\sum_{i=1}^\infty T_i\,Z_{T_i}(\beta_1(T_i)/\lambda)\,\,
\sigma(a-T_i)\,\,\overline \kappa_0^-(a-T_i)\,\chi_{T_i<a}.
\end{eqnarray*}
Note that the first term at the r.h.s.\ does not depend on the random variables $T_i$. Therefore, 
\begin{eqnarray*}
\overline	\kappa_0^-(a) 
&=& \widehat\kappa(a)\, 
e^{ -p\int_0^a(a-c)\beta_0(c)\,
\sigma(c)\overline \kappa_0^-(c)\,dc}\,\,\\
&& \qquad\qquad \times\quad E\bigg(\exp\bigg\{-p\,
\sum_{i=1}^\infty T_i\,Z_{T_i}(\beta_1(T_i)/\lambda)\,
\sigma(a-T_i)\overline \kappa_0^-(a-T_i)\,\chi_{T_i<a}\,\bigg\}\bigg).
\end{eqnarray*}
We focus on the expectation. 
Let ${\mathcal E}_k$, $k\in\N$, denote the event that $T_1,\ldots,T_{k-1}\in[0,a]\,\mbox{and }\,T_k\geq a$.
Recall that $Y_a$ is the counting process associated to $(T_i)_{i\in\N})$, s.t. 
$$P({\mathcal E}_k) = P(Y_a=k-1) =  \frac {(\lambda a)^{k-1}}  {(k-1)!} e^{-\lambda a}.$$
 With the notation of Lemma~\ref{lemmaPoissonRandom} we have 
\begin{eqnarray*}
\ast &:=&E\bigg(\exp\bigg\{-p
\sum_{i=1}^\infty T_i\,Z_{T_i}(\beta_1(T_i)/\lambda)\,
\sigma(a-T_i)\overline \kappa_0^-(a-T_i)\,\chi_{T_i<a}\,\bigg\}\bigg)\\
&=&	\sum_{k=1}^\infty 
E\bigg(\exp\bigg\{-p\,
\sum_{T\in  \Omega_{a,k-1}} T\,Z_{T}(\beta_1(T)/\lambda)\,
\sigma(a-T)\overline \kappa_0^-(a-T)\,\bigg\}
\,\bigg|\,{\mathcal E_k}\bigg)\,P({\mathcal E}_k).
\end{eqnarray*}
According to Lemma \ref{lemmaPoissonRandom}, $\Omega_{a,k}$ 
can be replaced by $\tilde \Omega_{a,k}$, which is the set of $k$ i.i.d.\ in $[0,a]$ uniformly distributed random variables. We proceed 
\begin{eqnarray*}
\ast &=&	\sum_{k=1}^\infty 
E\bigg(\exp\bigg\{-p\,
\sum_{T\in \tilde \Omega_{a,k-1}}  T\,Z_{T}(\beta_1(T)/\lambda)\,
\sigma(a-T)\overline \kappa_0^-(a-T)\,\bigg\}\bigg)\,
\frac 1 {(k-1)!}\,(\lambda a)^{k-1}e^{-\lambda a}\\
 &=&	\sum_{k=1}^\infty 
E\bigg(\exp\bigg\{-p\,
\sum_{T\in \tilde \Omega_{a,k-1}} (a-T)\,Z_{a-T}(\beta_1(a-T)/\lambda)\,
\sigma(T)\overline \kappa_0^-(T)\,\bigg\}\bigg)\,
\frac 1 {(k-1)!}\,(\lambda a)^{k-1}e^{-\lambda a}\\
&=&	\sum_{k=1}^\infty 
E\bigg(\prod_{T\in \tilde \Omega_{a,k-1}}\exp\bigg\{-p\,
 (a-T)\,Z_{a-T}(\beta_1(a-T)/\lambda)\,
\sigma(T)\overline \kappa_0^-(T)\,\bigg\}\bigg)\,
\frac 1 {(k-1)!}\,(\lambda a)^{k-1}e^{-\lambda a }\\
&=&	\sum_{k=1}^\infty \,E\bigg(\,
\prod_{T\in \tilde \Omega_{a,k-1}}E\bigg(\exp\bigg\{-p\,
(a-T)\,Z_{a-T}(\beta_1(a-T)/\lambda)\,
\sigma(T)\overline \kappa_0^-(T)\,\bigg\}\bigg)\,\bigg)\,
\frac 1 {(k-1)!}\,(\lambda a)^{k-1}e^{-\lambda a }\\
&=&	\sum_{k=1}^\infty 
\bigg\{
\frac 1 a \int_0^a E\bigg(\exp\bigg\{-p\,
(a-\tau)\,Z_{a-\tau}(\beta_1(a-\tau)/\lambda)\,
\sigma(\tau)\overline \kappa_0^-(\tau)\,\bigg\}\bigg)\,d\tau\,
\bigg\}^{k-1}
\frac 1 {(k-1)!}\,(\lambda a)^{k-1}e^{-\lambda a }
\end{eqnarray*}
where we used the pairwise independency of the random variables in $\tilde\Omega_{a,k-1}$. Let us consider the remaining expectation, 
\begin{eqnarray*}
&&E\bigg(\exp\bigg\{-p\,
(a-\tau)\,Z_{a-\tau}(\beta_1(a-\tau)/\lambda)\,
\sigma(\tau)\overline \kappa_0^-(\tau)\,\bigg\}\bigg)\\
&=& \sum_{\ell=0}^\infty 
\exp\bigg\{-p\,
(a-\tau)\,\ell\,
\sigma(\tau)\overline \kappa_0^-(\tau)\,\bigg\}\,\,
\frac 1 {\ell!}(\beta_1(a-\tau)/\lambda)^\ell e^{-\beta_1(\tau)/\lambda}\\
&=& 
\exp\bigg(\bigg[e^{-p(a-\tau)\sigma(\tau)\ol\kappa_0^-(\tau)}-1\bigg]\beta_1(a-\tau)/\lambda\bigg).
\end{eqnarray*}
Therewith,
\begin{eqnarray*}
\ast &=&	\sum_{k=0}^\infty 
\bigg(\frac 1 a\,\int_0^a\exp\bigg\{\bigg[e^{-p(a-\tau)\sigma(\tau)\ol\kappa_0^-(\tau)}-1\bigg]\beta_1(a-\tau)/\lambda\,\bigg\}\, d\tau\bigg)^k\,
\frac 1 {k!}\,(\lambda a)^ke^{-\lambda a }\\
&=&	
\exp\bigg(\lambda\int_0^a\exp\bigg\{\bigg[e^{-p(a-\tau)\sigma(\tau)\ol\kappa_0^-(\tau)}-1\bigg]\beta_1(a-\tau)/\lambda\,\bigg\}\, d\tau\bigg)\,e^{-\lambda a }\\
&=& W_{\lambda\,a}\bigg(\frac 1 a \int_0^a\exp\bigg\{\bigg[e^{-p(a-\tau)\sigma(\tau)\ol\kappa_0^-(\tau)}-1\bigg]\beta_1(a-\tau)/\lambda\,\bigg\}\, d\tau\bigg).
\end{eqnarray*}
\qed\par\medskip

%
%

\begin{rem}\label{backwLargeLambda}
	We consider the two extreme cases, $\lambda\rightarrow\infty$, and $\lambda\rightarrow 0$. We focus on one-step tracing, the computation/results for recursive tracing are parallel.\\
Defining $\eta(\tau,a)=\left[e^{-p(a-\tau)\sigma(\tau)\ol\kappa_0^-(\tau)}-1\right]\beta_1(a-\tau)$ yields
\begin{eqnarray*}
&&\lim_{\lambda\rightarrow\infty}
W_{\lambda\,a}\bigg(\frac 1 a \int_0^a\exp\bigg\{-\,\lambda^{-1}
\, \eta(\tau,a)\,\bigg\}\, d\tau\bigg)\\
&=& 
\lim_{\lambda\rightarrow\infty} 
\exp\bigg(\lambda\,\int_0^a\bigg[\exp\bigg\{-\,\lambda^{-1}
\,\eta(\tau,a)\,\bigg\}\, -1\bigg]\,\,\,d\tau\bigg) \\
&=& 
\lim_{\lambda\rightarrow\infty} 
\exp\bigg(\lambda\int_0^a-\,\lambda^{-1}
\,\eta(\tau,a)\,+{\cal O}(\lambda^{-2})\,\,\,d\tau\bigg) 
=  \exp\bigg(-\int_0^a
\,\eta(\tau,a)\,\,\,d\tau\bigg).
\end{eqnarray*}
In the limit, we therefore do not simply obtain the situation with a deterministic contact rate $\beta_0(a)+\beta_1(a)$. Intuitively, for $\lambda$ large, the very frequent and very small random contact events should behave like a deterministic contact rate. Interestingly, the additional variance in $Z_a(\beta(a)/\lambda)$ has an effect also in the limiting case. For $p$ small, however, we have that 
$$\eta(\tau,a)=\left[e^{-p(a-\tau)\sigma(\tau)\ol\kappa_0^-(\tau)}-1\right]\beta_1(a-\tau)\approx -p(a-\tau)\sigma(\tau)\ol\kappa_0^-(\tau)\,\beta_1(a-\tau),$$  
s.t.\ in case of $\lambda\rightarrow \infty$ and $p\ll 1$,
	\begin{eqnarray*}
	\overline	\kappa_0^-(a) 
	&\approx& \widehat\kappa(a)\, 
	e^{ -p\int_0^a(a-c)(\beta_0(a-c)+\beta_1(a-c))\,
		\sigma(c)\overline \kappa_0^-(c)\,dc}
	\end{eqnarray*}
and the random contacts behave as deterministic contacts.\par\medskip 

In the other extreme ($\lambda\rightarrow 0$) we find that 
$$
\lim_{\lambda\rightarrow 0} 
\exp\bigg(\lambda\,\int_0^a\bigg[\exp\bigg\{-\,\lambda^{-1}
\,\eta(\tau,a)\,\bigg\}\, -1\bigg]\,\,\,d\tau\bigg)= 1$$
such that 
\begin{eqnarray*}
\lim_{\lambda\rightarrow 0}	\overline	\kappa_0^-(a) 
&=& \widehat\kappa(a)\, 
e^{ -p\int_0^a(a-c)\beta_0(a-c)\,
	(\overline\kappa_0^-{'}(c)-\mu(c)\overline\kappa_0^-(c))\,dc}\,.
\end{eqnarray*}
That is, the rare (but large) spreading events modeled by the random contact rate do not play a role in the effect of backward tracing. 
If $\lambda$ is small, the fraction of super-spreaders among all infected individuals is negligible. In contrast, the fraction of their infectees is large. As each super-spreader produces a high number of infectees, he/she is detected and removed soon. However, as we only have very few super-spreaders, this fact hardly affects the probability to be infected at a.s.i.\ $a$ for the average infected person. This situation resembles the observation that rare super-spreader events hardly affect the probability for a major outbreak (see Remark~\ref{majorOurbreak}). 
\end{rem}

We aim at a computation of $R_{eff}$; later, we will also address the exponential growth rate. The difficulty here is that the time span a given individual is infectious is correlated with the infection rate -- an individual who produces many infectees by a superspreader event is likely to be discovered early. We need to address this correlation. The next computations and propositions serve this fact. Thereto, we define a function $H:\R\rightarrow\R$ 
$$ H(r) = E\bigg(\int_0^\infty \beta(a)\, \kappa^-_0(a)\, e^{-ra}\, da  \bigg).$$
Then, the reproduction number with CT is the integral over the contact rate times the survival probability $\kappa^-_0(a)$, $R_{eff}=H(0)$. In case of independent generations (which is not true for contact tracing) the exponential growth rate $r$ is given by the unique root of $H(r)=1$~\cite{Wallinga2006}. Nevertheless, we will later use also that formula to identify the exponential growth, though it is in our setting only a heuristic approach and not a hard result (see discussion below). 
We handle the deterministic component $\beta_0(a)$ and the random component $Z_a(\beta_1(a)/\lambda) dY_a$ of the contact rate separately. 
For the deterministic component, we simply find 
$$ E\bigg(\int_0^\infty \beta_0(a)\, \kappa^-_0(a)\, e^{-ra}\, da  \bigg) =  \int_0^\infty \beta_0(a)\, \ol\kappa^-_0(a)\, e^{-ra}\, da.$$
The expectation of the random component $\int_0^\infty Z_a(\beta_1(a)/\lambda)\kappa_0^-(a)\, dY_a$ is more subtle to determine, as in a realization the contact rate $Z_a(\beta_1(a)/\lambda) dY_a$ precisely is the contact rate of the focal individual which appears in the (random) function  $\kappa_0^-(a)$. This observation indicates dependencies which we need to take into account. 

\begin{prop}\label{backwardResultPropToComputeReff} Let $\varphi(a)$ a continuous function. Then, for one-step tracing, we have 
\begin{eqnarray*}
 E\bigg(\int_0^\infty \varphi(a) Z_a(\beta_1(a)/\lambda)\kappa_0^-(a)\, dY_a\bigg) = 
\int_0^\infty \beta_1(a)\,\varphi(a)\, \ol\kappa_0^-(a)\, da.
\end{eqnarray*}
\end{prop}
{\bf Proof: } If we spell out the integral 
$E\bigg(\int_0^\infty Z_a(\beta_1(a)/\lambda)\kappa_0^-(a)\, dY_a\bigg)$ we find 
$$
E\bigg(\int_0^\infty \varphi(a) Z_a(\beta_1(a)/\lambda))\,
\widehat \kappa(a)\,e^{ -p\int_0^a(a-c)\beta_0(a-c)\,
	\sigma(c)\overline \kappa_0^-(c)\,\,dc}\,e^{-p\int_0^a(a-c)\,Z_{a-c}(\beta_1(a-c)/\lambda)\sigma(c)\ol\kappa_0^-(c)\, dY_c}\, dY_a \bigg).
$$
Importantly, the two counting measures $Y_a$ and $Y_c$ belong to the same realization of the Poisson process, and so are the Poisson random variables  $Z_a(.)$ at identical arrival times of the Poisson process identical realizations. Therefore some attention is required in computing the expectation. The argument, however, resembles that used in the proof of  Theorem~\ref{backwardTheorem}. 
Let 
$$\tilde\varphi(a) =  \widehat\kappa(a)\, 
e^{ -p\int_0^a(a-c)\beta_0(a-c)\,
	\sigma(c)\overline \kappa_0^-(c)\,\,dc}\varphi(a).$$
Therewith, the integral we aim to simplify becomes 
\begin{eqnarray*}
&&E\bigg(\int_0^\infty \tilde\varphi(a) Z_a(\beta_1(a)/\lambda)\,
\,e^{-p\int_0^a(a-c)\,Z_{a-c}(\beta_1(a-c)/\lambda)\sigma(c)\ol\kappa_0^-(c)\, dY_c}\, dY_a\bigg)\\
&=&E\bigg(\int_0^\infty \tilde\varphi(a) Z_a(\beta_1(a)/\lambda)\,
\,e^{-p\int_0^ac\,Z_c(\beta_1(c)/\lambda)\sigma(a-c)\ol\kappa_0^-(a-c)\, dY_c}\, dY_a\bigg)\\
&=&
\sum_{i=1}^\infty 
E\bigg( \tilde\varphi(T_i) Z_{T_i}(\beta_1(T_i)/\lambda)\,
\,e^{-p\sum_{j=1}^{i-1}T_j\,Z_{T_j}(\beta_1(T_j)/\lambda)\sigma(T_i-T_j)\ol\kappa_0^-(T_i-T_j)\,}\, \bigg)\\
&=&
\sum_{i=1}^\infty \int_0^\infty 
E\bigg( \tilde\varphi(t) Z_{t}(\beta_1(t)/\lambda)\,
\,e^{-p\sum_{T_j\in\Omega_{t,i-1}}T_j\,Z_{T_j}(\beta_1(T_j)/\lambda)\sigma(t-T_j)\ol\kappa_0^-(t-T_j)\,}\, \bigg)\, \frac{\lambda\,(\lambda t)^{i-1}}{(i-1)!}\,e^{-\lambda t}\,\,dt\\
&=&
\sum_{i=1}^\infty \int_0^\infty 
E\bigg( \tilde\varphi(t) Z_{t}(\beta_1(t)/\lambda)\,
\,\prod_{j=1}^{i-1}\bigg[\frac 1 t\int_0^te^{-p\,s\,Z_{s}(\beta_1(s)/\lambda)\sigma(t-s))\ol\kappa_0^-(t-s)\,} \,ds\,\bigg]\bigg)\, \frac{\lambda\,(\lambda t)^{i-1}}{(i-1)!}\,e^{-\lambda t}\,\,dt\\
&=&
\sum_{i=1}^\infty \int_0^\infty 
E\bigg( \tilde\varphi(t) Z_{t}(\beta_1(t)/\lambda)\,
\,\prod_{j=1}^{i-1}\bigg[\frac 1 t\int_0^te^{-p(t-s)\,Z_{t-s}(\beta_1(t-s)/\lambda)\sigma(s))\ol\kappa_0^-(s)\,} \,ds\,\bigg]\bigg)\, \frac{\lambda\,(\lambda t)^{i-1}}{(i-1)!}\,e^{-\lambda t}\,\,dt\\
&=&
\sum_{i=1}^\infty 
\,\int_0^\infty \tilde\varphi(t)\,\beta_1(t)
\,\prod_{j=1}^{i-1}\bigg[\frac 1 t\int_0^t\sum_{k=0}^\infty e^{-p(t-s)\,k\sigma(s))\ol\kappa_0^-(s)\,}\,\frac{(\beta_1(t-s)/\lambda)^k}{k!}e^{-\beta_1(t-s)/\lambda} \,ds\bigg] \frac{(\lambda t)^{i-1}}{(i-1)!}\,e^{-\lambda t}\,\,dt\\
&=&
\sum_{i=1}^\infty 
\,\int_0^\infty \tilde\varphi(t)\,\beta_1(t)
\,\bigg[\frac 1 t\int_0^t
\exp\bigg\{\bigg[
e^{-p(t-s)\,\sigma(s))\ol\kappa_0^-(s)\,}-1\bigg]\,\beta_1(t-s)/\lambda
\bigg\}\,ds
\bigg]^{i-1} \frac{(\lambda t)^{i-1}}{(i-1)!}\,e^{-\lambda t}\,\,dt\\
&=&
\,\int_0^\infty \tilde\varphi(t)\,\beta_1(t)
\,W_{\lambda t} \bigg(
\frac 1 t \,\int_0^t
\exp\bigg\{\bigg[
e^{-p(t-s)\,\sigma(s))\ol\kappa_0^-(s)\,}-1\bigg]\,\beta_1(t-s)/\lambda
\bigg\}\,ds
\bigg) \,\,dt\\
&=& \int_0^\infty \varphi(t)\,\beta_1(t)\, \ol\kappa_0^-(t)\, dt.
\end{eqnarray*}
\qed\par\medskip 
For full tracing, the argument is exactly the same: In 
eqn.~(\ref{kappaMinusRekurs}), the functions $\ol\kappa_0^-(a)$ on the right hand side concern the infectees of the focal individual, s.t.\ this function  is independent of the focal individual's contact rate. We are allowed to replace 
$\sigma(c)\ol\kappa_0^-(c)$ by $-\ol\kappa_0^-(c)-\mu(c)\kappa_0^-(c)$, and use the same computations as above to find the next proposition.
\begin{prop} Let $\varphi(a)$ a continuous function. Then, for recursive  tracing, we have 
	\begin{eqnarray*}
		&&E\bigg(\int_0^\infty \varphi(a) Z_a(\beta_1(a)/\lambda))\,
		\widehat \kappa(a)\,e^{ -p\int_0^a(a-c)\beta_0(c)\,
				\,[-\ol\kappa_0^-(c)-\mu(c)\kappa_0^-(c))]\,\,dc}\,\times \\
		&&\qquad \times e^{-p\int_0^a(a-c)\,Z_c(\beta_1(c)/\lambda)
			\,[-\ol\kappa_0^-(c)-\mu(c)\kappa_0^-(c))]\, dY_c}\, dY_a \bigg) \\
		&=& 
		\int_0^\infty \beta_1(a)\,\varphi(a)\, \ol\kappa_0^-(a)\, da.
	\end{eqnarray*}
\end{prop}

With this understanding, we have the following corollary. 

\begin{cor} \label{backwRnullProp}For recursive as well as for one-step tracing, we find for $r\in\R$ that 
\begin{eqnarray} 
E\bigg(\int_0^\infty Z_a\left(\frac{\beta_1(a)}{\lambda}\right)\,E(\kappa_0^-(a)|\beta)\,e^{-ra}\,dY_a\bigg)
&=& 
\int_0^\infty \beta_1(a)\widehat\kappa(a)\, e^{-ra}\, 
\ol\kappa_0^-(a)\, da.
\end{eqnarray}
\end{cor}

From this corollary, we obtain 
\begin{theo} \label{backwRnullTheo}For one-step as well as for recursive tracing, we have 
$$R_{eff} = \int_0^\infty (\beta_0(a)+\beta_1(a))\,\ol\kappa_0^-(a)\, da.$$
\end{theo}

\begin{rem} It is remarkable that the formula for the effective 
reproduction number is identical for one-step and recursive tracing, were  
we emphasize that $\ol \kappa_0^-(a)$ does depend on the tracing mode. 
In that, there is an implicit influence of one step/recursive tracing. 
In any case, the intuition is that an individual is infectious for a 
random time span that is described by the ``survival probability'' 
$\kappa_0^-(a)$. During this time period, the individual produces infectees 
at rate $\beta_0(a)+\beta_1(a)$, in both modes of tracing. The correlation of the infectious period and the contact rate does not affect the equation. It is very well possible that the choice of the distributions in $\beta$ are crucial, and that dependencies will affect this formula if we go away from a Poisson process for the arrival times of superspreader events, or from a Poisson distribution for the size of superspreader events.
\end{rem}

\subsection{Forward tracing}
The investigation of forward tracing turns out to be more simple, as the contact rate of a focal individual does only influence the infectees, but not the infector. In that, the tracing probability is independent of the focal individual's contact rate, and taking expectations is rather straightforward.

The number of ancestors in the infection lineage affects a focal individual. Therefore, the generation matters; generation refers to the location of an individual in the tree (forest) of infection: The primary infected person has generation $0$, the secondary cases generation $1$, the infectees of the secondary infecteds generation $3$, etc.  Let $\ol\kappa_i^+(a)$ denote the probability of a generation $i$ individual to be infectious at a.s.i.\  $a$, and $\ol\kappa_i^+(a|b)$ that probability, conditioned on the age of the infector $b$ at the infection event.

\begin{prop}  \label{coditionedForwardRecursive} In case of recursive tracing, we have 
	for $i>0$
	\begin{eqnarray}
	\hspace*{-0.5cm}\ol\kappa^+_i(a|b) 
	&=&
	\frac{\widehat\kappa(a)}{\ol\kappa_{i-1}^+(b)}\,\, \bigg\{
	\ol\kappa_{i-1}^+(b) - p\,\int_0^{a} 
	\bigg(-(\ol\kappa_{i-1}^+(b+c))'\,-\,\mu(b+c)\, 
	\,\,\ol\kappa_{i-1}^+(b+c)\bigg)
	\,dc
	\bigg\}.
	\end{eqnarray}	
\end{prop}
{\bf Proof: }
Our focal individual is infectious if it did not recover independently of CT, times the probability that no tracing event did remove the individual from the class of infecteds, 
\begin{eqnarray*}
	\hspace*{-0.5cm}\ol\kappa_i^+(a|b) 
	&=&
	\widehat\kappa(a) \bigg\{
	1 - \,P(\mbox{a successful tracing event did happen}) 
	\bigg\}.
\end{eqnarray*}	
In order to obtain the probability for a successful tracing event, we first note that we know that the infector has been infectious at (his/her) a.s.i.\  $b$, s.t.\ the probability for him/her to be infectious at a.s.i.\  $b+c$ reads
$$ \frac{\ol\kappa_{i-1}^+(b+c)}{\ol\kappa_{i-1}^+(b)}.
$$
We use here the expected values (averaged over all possible realizations of $\beta$). 
The detection rate is the hazard rate minus the rate to recover spontaneously/unobserved, 
$$\frac{{-\ol\kappa_{i-1}^+(b+c)}'}{\ol\kappa_{i-1}^+(b+c)}-\mu(b+c).$$
Hence, the desired probability reads
$$ \mbox{a successful tracing event did happen} = p\,\int_{0}^{a} 
\bigg(\frac{{-\ol\kappa_{i-1}^+(b+c)}'}{\ol\kappa_{i-1}^+(b+c)}-\mu(b+c)\bigg)
\,\,\frac{\ol\kappa_{i-1}^+(b+c)}{\ol\kappa_{i-1}^+(b)}
\,dc.$$
\qed\par\medskip

\begin{prop} \label{coditionedForwardOneStep} In case of one-step-tracing we have 
	\begin{eqnarray}\label{coditionedForwardOneStepEqn1}
	\hspace*{-0.5cm}\ol \kappa_{i}^+(a|b) 
	&=&
	\widehat\kappa(a) \bigg\{
	1 - p\,\int_{0}^{a} 
	\sigma(b+c)
	\,\,\frac{\ol\kappa_{i-1}^+(b+c)}{\ol\kappa_{i-1}^+(b)}
	\,dc
	\bigg\}.
	\end{eqnarray}	
\end{prop}
{\bf Proof: } The argument parallels that of Proposition~\ref{coditionedForwardRecursive}. We only need to take into account that in one-step tracing the infector has to be detected directly, what happens at rate $\sigma(.)$. This rate replaces the hazard rate minus the spontaneous recovery rate.
\qed\par\medskip 

In order to determine the desired probability  $\ol \kappa^+_{i}(a)$ we remove the condition in $\ol\kappa_{i}^+(a|b)$. Thereto we determine the probability density for an infector to have age $b$ of infection. The net infection rate is $E(\beta(b)\,\kappa_{i-1}(b))$, where the $\beta$ is the contact rate of the $i-1$'th generation individual, that is independent of $\kappa_{i-1}$. Therefore, 
 $E(\beta(b)\,\kappa_{i-1}(b)) = (\beta_0(b)+\beta_1(b))\,\ol\kappa_{i-1}(b)$. 
 We normalize this expression  and find for the probability distribution of the age of the infector at infection events
$$\varphi_{i-1}(b) = \frac{(\beta_0(b)+\beta_1(b))\,\ol\kappa_{i-1}^+(b)}
{\int_0^\infty (\beta_0(c)+\beta_1(c))\,\ol\kappa_{i-1}^+(c)\, dc}.
$$

\begin{cor} \label{forwardCor} In one-step tracing as well as in recursive tracing, we have for $i>0$
	\begin{eqnarray}\label{coditionedForwardOneStepEqn2}
	\ol \kappa^+_{i}(a) &=& \int_0^\infty  \ol\kappa_{i}^+(a|b)\varphi_{i-1}(b)\, db.
	\end{eqnarray}
\end{cor}
\qed\par\medskip 

\begin{rem}
In contrast to backward tracing, forward tracing does not depend at all on $\lambda$ (and therewith on the dispersion factor). That's because each individual has a single infector, and also super-spreader events do not change that fact. 
\end{rem}
Because in backward tracing the randomness does not play a role, we immediately conclude the following result for the effective reproduction number.
\begin{cor} \label{forwRnullTheo}The effective reproduction number of an individual in generation $i$, $R_{eff,i}$, is given by 
	$$R_{eff,i} = \int_0^\infty (\beta_0(a)+\beta_1(a))\,\ol\kappa_i^+(a)\, da.$$
\end{cor}
 It is rather involving to obtain a proof for the convergence of $R_{eff,i}$ for $i\rightarrow\infty$. It is possible to show convergence of $(\kappa_i(a))_{i\in\N_0}$ in a weighted $L^1$ space if  $p$ is sufficiently small (no super-spreader events)~\cite{mueller2007}. This convergence is a hint that also $R_{eff,i}$ converges; numerical analysis seems to indicate that the restriction on $p$ is not necessary for the convergence. However, this problem is out of scope of the present paper.

\subsection{Full tracing}

Full tracing is the combination of backward- and forward tracing. The basic argument is as follows: 
We focus on an individual in generation $i$. As long as the individual is infectious, the infectees (generation $i+1$) cannot be traced by forward tracing (and thus are only subject to backward tracing), and the infector (generation $i-1$) cannot be traced by backward tracing triggered by the focal individual. The infectees and the infector of the focal individual decouple. As long as no forward tracing takes place, the focal individual is subject to backward tracing only. Backward tracing already is analyzed. The investigation that requires attention is forward tracing. It turns out that we cannot directly use our results for forward tracing to understand full tracing. We need to adapt the analysis at that point.\par\medskip

A central ingredient in the analysis of forward tracing above was the fact that the probability for the focal individuals infector to be infectious at a.s.i.\ $a+b$, if the focal individual has a.s.i.\ $a$, is given by  $\ol\kappa^+_{i-1}(a+b)/\ol\kappa^+_{i-1}(b)$. \\
Let us now think about forward tracing, in combination with backward tracing. The focal individual can be produced by a deterministic or a random (super-spreader) contact event. Recall that we assume that the focal infectee did not trigger a tracing event in age interval $[0,a]$. If the focal individual has been produced in a deterministic contact event, the knowledge that this individual exists does not change the structure of the infector's other infectees (number and timing), s.t.\ the infectors probability to be infectious at a.s.i. $a$ again is  $\ol\kappa_{i-1}(a+b)/\ol\kappa_{i-1}(b)$. This is different if a random tracing event did infect our focal individual: In this case, it is likely that many sibling infectees have been produced, which decease the infector's probability to be infectious, even though the target individual still is infectious.\\
The next interesting question is the following: Consider two individuals of generation $i-1$, both did trigger a random contact event at their a.s.i.\ $b$. One is the infector of our focal individual (that is, this infector did produce at least one infectee in the event), for the other one, we have no additional information about the number of infectees produced in the event. Is the probability to be infectious at a.s.i.\ $a+b$ for these two infectors different? That is, is the statistics of the number of infectees for the second infector similar to statistics of the number of siblings of our focal individual? Astonishingly, these two statistics coincide, indeed. This fact is known as the the ``environmental equivalence property'' in the context of Poisson games, and has been proven by Myerson~\cite{Myerson1998}. Myerson even shows that the Poisson distribution is (under mild conditions) the only distribution with that property. The assumption that the number of infectees in a super-spreader event is Poissonian distributed prevents our considerations to become technically highly involving. In that, this assumption is crucial.  \par\medskip

Let $\ol K_{i-1}(a+b|b,D)$ denote the probability for the infector to be infectious at a.s.i. $a+b$, if he/she produced 
the focal individual at a.s.i.\ $b$ in a deterministic contact event, $\ol K_{i-1}(a+b|b,R)$ this probability in case of a random contact event, and 
$\ol K_{i-1}(a+b|b)$ if nothing is known about the contact.

\begin{prop}\label{condProbabFull}  For recursive tracing, we have 
	\begin{eqnarray}
	\ol K_{i-1}(a+b|b,D) & = &  \frac{\ol\kappa_{i-1}(a+b)}{\ol\kappa_{i-1}(b)},\\
	\ol K_{i-1}(a+b|b,R) & = &   \frac{\ol\kappa_{i-1}(a+b)}{\ol\kappa_{i-1}(b)}\,\,\exp \bigg(-p\,(\beta_1(b)/\lambda)\,\int_{0}^{a} 
	\bigg(\frac{{-\ol\kappa_{0}(c)}'}{\ol\kappa_{0}(c)}-\mu(c)\bigg)
	\,\,\ol\kappa_{0}(c)
	\,dc\bigg),
	\end{eqnarray}
	while the result for one-step-tracing is given by 
		\begin{eqnarray}
	\ol K_{i-1}(a+b|b,D) & = &  \frac{\ol\kappa_{i-1}(a+b)}{\ol\kappa_{i-1}(b)},\\
	\ol K_{i-1}(a+b|b,R) & = &   \frac{\ol\kappa_{i-1}(a+b)}{\ol\kappa_{i-1}(b)}\,\,\exp \bigg(-p\,(\beta_1(b)/\lambda)\,\int_{0}^{a} 
	\sigma(b+c)\,
	\,\,\ol\kappa_{0}(c)
	\,dc\bigg).
	\end{eqnarray}
\end{prop}
{\bf Proof: } As a deterministic contact has no additional information that changes the probability to be infectious, we directly find 
$$\ol K_{i-1}(a+b|b,D) = \frac{\ol\kappa_{i-1}(a+b)}{\ol\kappa_{i-1}(b)}.$$
That is different in a random contact event. If the individual is not traced by the infectees generated in that event, we again have 
$ \frac{\ol\kappa_{i-1}(a+b)}{\ol\kappa_{i-1}(b)}$ as survival probability. 
The probability for a tracing event triggered by a given infectee (infected in that random contact event) is given by 
\begin{eqnarray*}
	p\,\int_{0}^{a} 
	\bigg(\frac{{-\ol\kappa_{0}(c)}'}{\ol\kappa_{0}(c)}-\mu(c)\bigg)
	\,\,\ol\kappa_{0}(c)
	\,dc.
\end{eqnarray*}
As the number of infectees is Poisson distributed with parameter $\beta_1(b)/\lambda$, we have for the probability that no infectee triggers successfully a tracing event
\begin{eqnarray*}
	&&	\sum_{k=0}^\infty \bigg(1-p\,\int_{0}^{a} 
	\bigg(\frac{{-\ol\kappa_{0}(c)}'}{\ol\kappa_{0}(c)}-\mu(c)\bigg)
	\,\,\ol\kappa_{0}(c)
	\,dc\bigg)^k \frac 1 {k!}(\beta_1(b)/\lambda)^k e^{-\beta_1(b)/\lambda}\\
	&=& \exp \bigg(-p\,(\beta_1(b)/\lambda)\,\int_{0}^{a} 
	\bigg(\frac{{-\ol\kappa_{0}(c)}'}{\ol\kappa_{0}(c)}-\mu(c)\bigg)
	\,\,\ol\kappa_{0}(c)
	\,dc\bigg).
\end{eqnarray*}
Hence, 
$$ \ol K_{i-1}(a+b|b,R)
=
\ol K_{i-1}(a+b|b,D)\,\exp \bigg(-p\,(\beta_1(b)/\lambda)\,\int_{0}^{a} 
\bigg(\frac{{-\ol\kappa_{0}(c)}'}{\ol\kappa_{0}(c)}-\mu(c)\bigg)
\,\,\ol\kappa_{0}(c)
\,dc\bigg).
$$
The result for one-step tracing follows by a parallel argument.
\qed\par\medskip 

The probability for the focal individual to be infected by a deterministic contact event is (we know that the infectors a.s.i.\ was $b$ at the time of infection)  $\beta_0(b)/(\beta_0(b)+\beta_1(b))$, and that the probability for a random contact event is $1-\beta_0(b)/(\beta_0(b)+\beta_1(b))$.
Hence we have the following corollary. 

\begin{cor} For recursive tracing, we have 
\begin{eqnarray}\label{fullProbabInfectorRekurs}
\ol K_{i-1}(a+b|b) &=&  \frac{\ol\kappa_{i-1}(a+b)}{\ol\kappa_{i-1}(b)} \,\,\times\\
&& \times \bigg(\,
1- \frac{\beta_1(b)}{\beta_0(b)+\beta_1(b)} 
 \,\,\bigg(1-\exp \bigg(-p\,(\beta_1(b)/\lambda)\,\int_{0}^{a} 
\bigg(\frac{{-\ol\kappa_{0}(c)}'}{\ol\kappa_{0}(c)}-\mu(c)\bigg)
\,\,\ol\kappa_{0}(c)
\,dc\bigg)\,\bigg)\,\bigg),\nonumber 
\end{eqnarray}
and for one-step-tracing, 
\begin{eqnarray}\label{fullProbabInfectorOneStep}
\ol K_{i-1}(a+b|b) &=&  \frac{\ol\kappa_{i-1}(a+b)}{\ol\kappa_{i-1}(b)}\times\\
&& \times \bigg(
1- \frac{\beta_1(b)}{\beta_0(b)+\beta_1(b)} \,\, 
\,\,\bigg(1-\exp \bigg(-p\,(\beta_1(b)/\lambda)\,\int_{0}^{a} 
\sigma(b+c)\,
\,\,\ol\kappa_{0}(c)
\,dc\bigg)\,\bigg)\,\bigg).\nonumber
\end{eqnarray}
\end{cor}

We now proceed to the probability 
$\kappa_i(a|b)$ for the infectee: the probability to be infectious at a.s.i.\ $a$, if the infector has had a.s.i.\ $b$ at the infection event. With the argument of Proposition~\ref{coditionedForwardRecursive}, where  we take into account that the conditioned probability for the infector is $K_{i-1}(a+b|b)$, instead of $\ol \kappa_{i-1}(a+b)/\ol \kappa_{i-1}(b)$, we find the following corollary.

\begin{prop} For  recursive tracing, we find 
\begin{eqnarray}\label{probabFokalConditionedFullRekurs}
\ol \kappa_i(a|b)
&=&   \ol\kappa_0(a)\,\,\,\bigg(1-p\,\int_{0}^{a} 
\bigg(\frac{-\ol K_{i-1}'(b+c|b)}{\ol K_{i-1}(b+c|b)}-\mu(b+c)\bigg)
\,\,\ol K_{i-1}(b+c|b)
\,dc\bigg)\nonumber\\
&=& 
 \ol\kappa_0(a)\,\,\,\bigg[1-p\,
\bigg( 1-\ol K_{i-1}(a+b|b)
-\int_{0}^{a}\mu(b+c)
\,\,\ol K_{i-1}(b+c|b)
\,dc\bigg)\bigg]
\end{eqnarray}
while for one-step tracing, we have
\begin{eqnarray}\label{probabFokalConditionedFullOneStep}
\ol \kappa_i(a|b)
=   \ol\kappa_0(a)\bigg(1-p\,\int_{0}^{a} 
\,\,\sigma(b+c)\,\,
\,\,\ol K_{i-1}(b+c|b)dc\bigg). 
\end{eqnarray}
\end{prop}

The last step necessary to complete the step from generation $i-1$ to generation $i$ is to remove the condition on the age of the infector. Thereto we use again that the age-of-infection distribution of the infector is given by
$$\varphi_{i-1}(b) = \frac{(\beta_0(b)+\beta_1(b))\,\ol\kappa_{i-1}(b)}
{\int_0^\infty (\beta_0(c)+\beta_1(c))\,\ol\kappa_{i-1}(c)\, dc}.
$$

We collect all results, and wrap up the induction step in the next two theorems, where we introduce $\hat K_{i-1}(a|b) = \ol K_{i-1}(a+b|b)\, \ol\kappa_{i-1}(b)$. We start with the remark that the survival probability is given by that of backward tracing (eqn.~(\ref{backExpectRec}) and (\ref{backExpectOneStep}); we drop now the minus in the index), multiplied by the probability to be not removed by forward tracing. In order to construct this second probability, we require the probability for the infector to be infectious ($\hat K_i(a|b)= \ol K_{i-1}(a+b|b)\, \ol\kappa_{i-1}(b)$ resp.\ $\tilde K_i(a+b|b)$ given in eqn.~(\ref{fullProbabInfectorRekurs}) and (\ref{fullProbabInfectorOneStep})), which then enters in the probability of a focal individual to be infectious at age $a$ since infection, given the infector has had age $b$ at the infectious event (eqn.~\ref{probabFokalConditionedFullRekurs}) and (\ref{probabFokalConditionedFullOneStep})). Last, the condition on $b$ is removed in integrating over the distribution of the age-since-infection at infectious events.

\begin{theo} Consider recursive tracing. 
	Let  (for $i>0$)
	\begin{eqnarray}
	\overline	\kappa_0(a) 
	&=& \widehat\kappa(a)\, 
	e^{ -p\int_0^a(a-c)\beta_0(a-c)\,
		\left(\frac{\overline\kappa_0{'}(c)}{\overline\kappa_0(c)}-\mu(c)\right)\overline\kappa_0(c)\,dc}\,\,\times  \\
	&&
	\quad\times  W_{\lambda\,a}\bigg(\frac 1 a \int_0^a\exp\bigg\{\bigg[e^{-p\,
		(a-c)\,
		(-\overline\kappa_0^-{'}(c)-\mu(c)\overline\kappa_0^-(c))}
	-1\bigg]
	\,\beta_1(a-c)\lambda^{-1}
	\,\bigg\}\, dc\bigg),\nonumber\\
\hat K_{i-1}(a|b)\, 
&=&  \ol\kappa_{i-1}(a+b) \,\,\times\\
&& \times \bigg(\,
1- \frac{\beta_1(b)}{\beta_0(b)+\beta_1(b)} 
\,\,\bigg(1-\exp \bigg(-p\,(\beta_1(b)/\lambda)\,\int_{0}^{a} 
\bigg(\frac{{-\ol\kappa_{0}{'}(c)}}{\ol\kappa_{0}(c)}-\mu(c)\bigg)
\,\,\ol\kappa_{0}(c)
\,dc\bigg)\,\bigg)\,\bigg),\nonumber\\
	\ol \kappa_i(a|b)\ol\kappa_{i-1}(b)
&=&   \ol\kappa_0(a)\,\,\,\bigg[(1-p)\ol\kappa_{i-1}(b)+p\,
\hat  K_{i-1}(a|b)
+p\,\int_{0}^{a}\mu(b+c)
\,\,\hat K_{i-1}(c|b)
\,dc\bigg)\bigg].\label{xeqnUseLater}
	\end{eqnarray}
	Then, 
	\begin{eqnarray*}
	\ol \kappa_{i}(a) &=& 
	\frac{\int_0^\infty  \ol\kappa_{i}(a|b)\,\ol\kappa_{i-1}(b)\,(\beta_0(b)+\beta_1(b))\, db}
	{\int_0^\infty (\beta_0(c)+\beta_1(c))\,\ol\kappa_{i-1}(c)\, dc}.
	\end{eqnarray*}
\end{theo}

\begin{figure}[htbp]
	\begin{center}
		(a)\includegraphics[width=4.4cm]{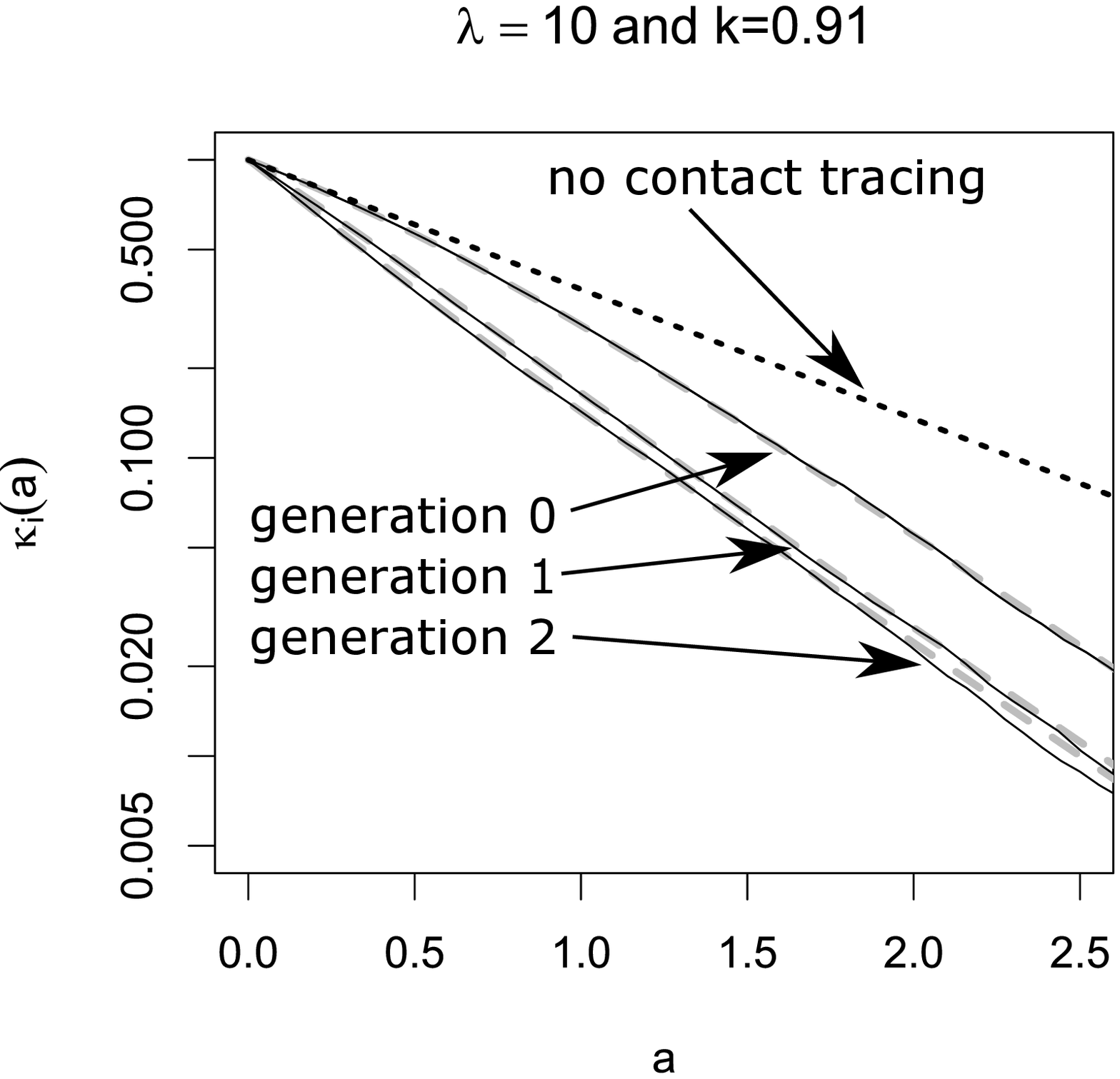}
		(b)\includegraphics[width=4.4cm]{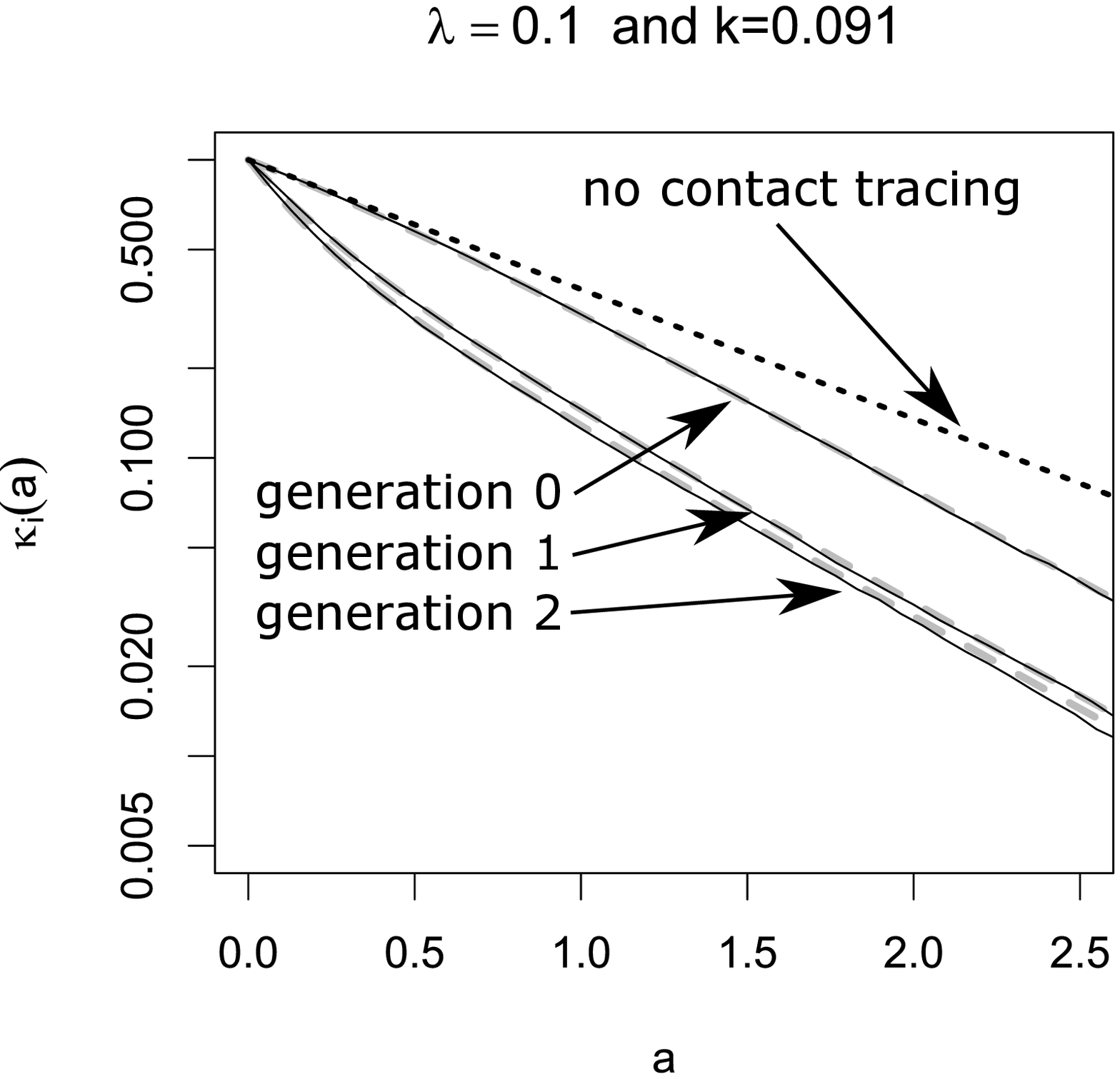}
		(c)\includegraphics[width=4.4cm]{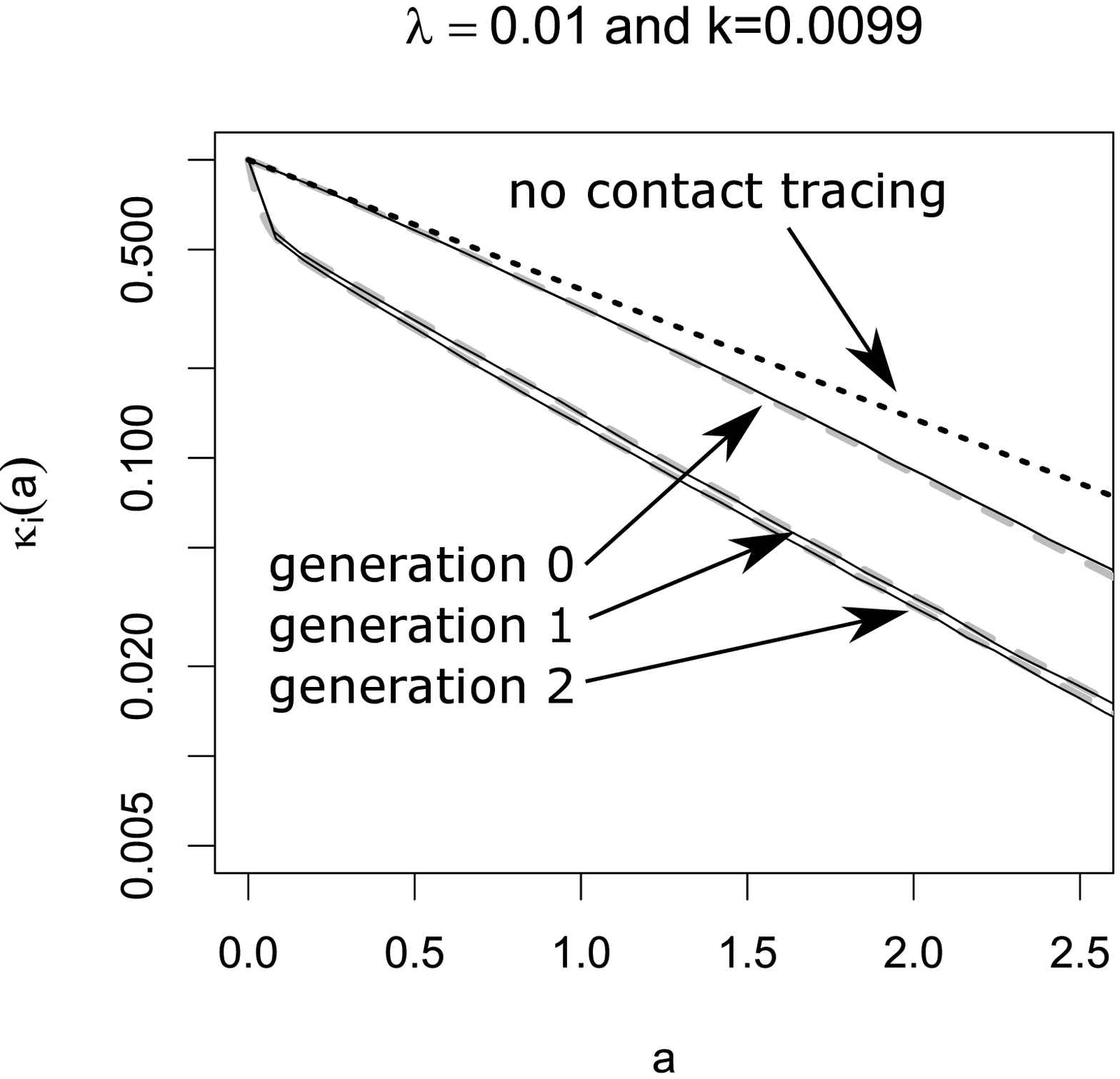}
	\end{center}
	\caption{Comparison of theory (gray dashed) and simulations (solid line) for generation~0, generation~1, and generation~2; dotted black line: $\widehat\kappa(a)$. Note the logarithmic scale in the y-axis. Parameter: $ \beta_0=\beta_1=0.75$, $\mu=\sigma=0.5$, $p=0.8$; recursive full tracing; $\lambda$ as indicated.}
	\label{simulTheo}
\end{figure}

\begin{theo}\label{fullOneStep} Consider one-step tracing. 
	Let (for $i>0$)
	\begin{eqnarray}
	\overline	\kappa_0(a) 
	&=& \widehat\kappa(a)\, 
	e^{ -p\int_0^a(a-c)\beta_0(a-c)\,
		\sigma(c)\,\overline\kappa_0(c)\,dc}\,\, \\
	&&
	\times  W_{\lambda\,a}\bigg(\frac 1 a \int_0^a\exp\bigg\{\bigg[e^{-p(a-c)\sigma(c)\ol\kappa_0^-(c)}-1\bigg]\beta_1(a-c)/\lambda\,\bigg\}\, dc\bigg)\nonumber\\
\ol K_{i-1}(a+b|b)& = & \frac{\ol\kappa_{i-1}(a+b)}{\ol\kappa_{i-1}(b)} \bigg(
1- \frac{\beta_1(b)}{\beta_0(b)+\beta_1(b)} \,\, 
\,\,\bigg(1-\exp \bigg(-p\,(\beta_1(b)/\lambda)\,\int_{0}^{a} 
\sigma(b+c)\,
\,\,\ol\kappa_{0}(c)
\,dc\bigg)\,\bigg)\,\bigg)\nonumber\\
	\ol \kappa_i(a|b)
	&=&   \ol\kappa_0(a)\bigg(1-p\,\int_{0}^{a} 
	\,\,\sigma(b+c)\,\,
	\,\,\ol K_{i-1}(b+c|b)dc\bigg)\label{yeqnUseLater}
\end{eqnarray}
Then, 
\begin{eqnarray*}
	\ol \kappa_{i}(a) &=& 
\frac{\int_0^\infty  \ol\kappa_{i}(a|b)\,\ol\kappa_{i-1}(b)\,(\beta_0(b)+\beta_1(b))\, db}
{\int_0^\infty (\beta_0(c)+\beta_1(c))\,\ol\kappa_{i-1}(c)\, dc}.
\end{eqnarray*}
\end{theo}

Again, the effective reproduction number again is an interesting quantity. 
Thereto we repeat the arguments for theorem~\ref{backwRnullTheo} in the 
present setting. We start with the following technical proposition that 
parallels proposition~\ref{backwardResultPropToComputeReff}.  
\begin{prop} Let $\varphi(a)$ be a bounded continuous function. Then, for one-step 
tracing, we have 
\begin{eqnarray*}
	E\bigg(	\int_0^\infty \varphi(a) Z_a(\beta_1(a)/\lambda)\kappa_i(a)\, dY_a\bigg) = 
		\int_0^\infty \beta_1(a)\,\varphi(a)\, \ol\kappa_i(a)\, da.
\end{eqnarray*}
\end{prop}
{\bf Proof: } First we obtain the equation for $\kappa_i(a)$. Recall that 
we fix the realization of the focal's individual contact rate, but take the 
average over the infector's and infectee's contact rates. Therefore, 
equation (\ref{fullProbabInfectorOneStep}) is still appropriate: 
Even if the focal individual has been infected by a random event, the functions 
$\ol K_{i-1}(a+b|b,D)$ and $\ol K_{i-1}(a+b|b,R)$ only depend on 
the siblings of our focal individual. In equation (\ref{probabFokalConditionedFullOneStep}), however, $\kappa_0(a)$ is dependent on the focal individual's contact rate. That is, 
$$\kappa_i(a|b) = \kappa_0(a)\,
\bigg(1-p\,\int_{0}^{a} 
\,\,\sigma(b+c)\,\,
\,\,\ol K_{i-1}(b+c|b)dc\bigg).$$
Therewith, we find 
\begin{eqnarray*}
&&	E\bigg(	\int_0^\infty \varphi(a) Z_a(\beta_1(a)/\lambda)\kappa_i(a)\, dY_a\bigg) \\
&=& 
	E\bigg(	\int_0^\infty \varphi(a) Z_a(\beta_1(a)/\lambda)	\frac{\int_0^\infty  \ol\kappa_{i}(a|b)\,\kappa_{i-1}(b)\,(\beta_0(b)+\beta_1(b))\, db}
	{\int_0^\infty (\beta_0(c)+\beta_1(c))\,\ol\kappa_{i-1}(c)\, dc}\,da\bigg)\\
&=& \frac{
\int_0^\infty E\bigg(\int_0^\infty \varphi(a) Z_a(\beta_1(a)/\lambda)	 \kappa_{i}(a|b)\,da\,\bigg)\,\ol\kappa_{i-1}(b)\,(\beta_0(b)+\beta_1(b))\, db}
{\int_0^\infty (\beta_0(c)+\beta_1(c))\,\ol\kappa_{i-1}(c)\, dc}\bigg)\\
\end{eqnarray*}
If we lump the product of $\varphi(a)$ and $\bigg(1-p\,\int_{0}^{a} 
\,\,\sigma(b+c)\,\,
\,\,\ol K_{i-1}(b+c|b)dc\bigg)$ into a single continuous function $g(a;b)$, the expectation reads $E\bigg(\int_0^\infty g(a;b)Z_a(\beta_1(a)/\lambda)	 \kappa_{0}(a)\,da\,\bigg)$. As we know that $\kappa_0(a)$ is identical with $\kappa_0^-(a)$ (backward tracing), we use proposition~\ref{probabFokalConditionedFullOneStep} to conclude 
$$E\bigg(\int_0^\infty g(a;b)Z_a(\beta_1(a)/\lambda)	 \kappa_{0}(a)\,da\,\bigg)
=
\int_0^\infty g(a;b)\beta_1(a) \ol\kappa_{0}(a)\,da.
$$
This formula implies the desired result.\par\qed\par\medskip 
As before, the proposition also holds true for recursive tracing. 
This proposition allows to compute the effective reproduction number for an individual of generation $i$. 
\begin{theo} \label{backwRnullTheo} For one-step as well as for recursive tracing, we have 
	$$R_{eff, i} = \int_0^\infty (\beta_0(a)+\beta_1(a))\,\ol\kappa_i(a)\, da.$$
\end{theo}
As in the case of backward tracing, we expect that the convergence of $R_{eff,i}$ for $i\rightarrow\infty$ is difficult to prove, though numerically we fund that this convergence is rapid. \par\medskip

The comparison of our theory with Monte-Carlo simulations can be found in Fig.~\ref{simulTheo}. 
The behavior for $\lambda\rightarrow 0$ is interesting. In that case, super-spreader events are extremely rare, but also extremely large.  Let us consider really the extreme case, in that not only $\lambda\approx 0$, but also $\beta_0\approx 0$, s.t.\ the infection is only driven by random contact events. In the limiting case, $\ol K_{i-1}(a+b|b)$ jumps from $1$ at $b=0$ to zero for $b>0$. Therefore, in one-step-tracing (see equation~(\ref{yeqnUseLater})) we have $ \kappa_{i+1}(a) = \widehat\kappa(a)$, while we obtain in recursive tracing  (using equation (\ref{xeqnUseLater})) that
$$ \kappa_{i+1}(a) = (1-p)\widehat\kappa(a)$$
 for $a>0$.
The interpretation is clear: Super-spreaders are a.s.\ rapidly detected and removed, but the fraction of super-spreaders within all infecteds is negligible small. Therefore, one-step tracing has no effect at all. In recursive tracing, on the other  hand, each infectee has the probability $p$ to be rapidly removed, and the probability to be infectious is decreased by the probability to escape CT. Hence we obtain in this limiting case
\begin{eqnarray}
 R_{eff}=(1-p)R_0,\label{ReffSimle}
 \end{eqnarray}
as already stated in~\cite{eames2003contact}. The critical tracing probability for which $R_{eff}=1$  is hence $p_{crit}=1-1/R_0$. That formula coincides with the critical tracing probability for (constant coefficients and) $\mu=0$, that is, in the scenario that all infecteds eventually develop symptoms~\cite{mueller2000}.

\begin{figure}[t!]
	\begin{center}
		\includegraphics[width=\textwidth]{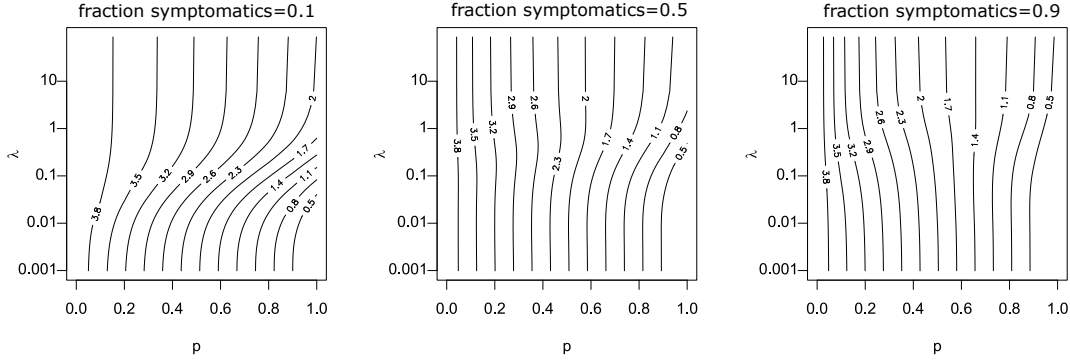}
	\end{center}
	\caption{Scan of $R_{eff}$ for different fractions of symptomatic cases. Please note the logarithmic scale on the $y$-axis (recursive tracing with $\beta_0=0$, $\beta_1=4$, $\mu=1-\sigma$, $\sigma=$ fraction of symptomatic cases).}\label{R0Scan}
\end{figure}

\subsection{Implications for the tracing-branching process}

As discussed above,  $R_{eff,i}$, and also the exponential growth rate $r_i$ given by the root of the equation
\begin{eqnarray}
1 = H_i(r) 
  =  \int_0^\infty (\beta_0(a)+\beta_1(a))\,\ol\kappa_i(a)\, e^{-r\,a}\, da 
\end{eqnarray}
is determined via the marginal ``survival'' probability $\ol\kappa_i(a)$. The function $H_i(r)$ (and in that also  $R_{eff,i}=H_i(0)$ and $r_i$) does take into account the correlation of an individuals survival rate and its contact rate, but is averaged over all other individuals. Due to the correlations between individuals of different generations in the tracing-branching process, the significance of $R_{eff}$ and the asymptotic growth rate  $\lim_{i\rightarrow\infty} r_i$  (if the limit exists) is not clear.\par\medskip 

We conjecture that these two quantities resemble the parallel quantities in the branching process theory, as the  correlations introduced by CT are strongly localized: We can couple the branching process ($p=0$) with the branching-tracing process ($p>0$). 
We start with the branching process, s.t.\ individuals are present until they are removed at rate $\mu(a)+\sigma(a)$. During their ``live span'', individuals infect contactees at the random rate $\beta(a)$. We introduce (as usual) two colors: living individuals and ghosts. If a tracing event leads to the discovery of a person (which induces the removal of that individual in the branching-tracing process), we change the color of that individual from ``living'' to ``ghost''. Ghosts cannot trigger further tracing events. Furthermore, the infectees of a living individual are living individuals, and the infectees of ghosts are ghosts. We start the process with one single living individual. \\
This coupling shows that the connected components of living individuals in the branching process are stochastically larger than that of the branching-tracing process. For constant parameters, in~\cite{mueller2003} the statistics of the connected components in the branching process has been investigated: In the long run, that size is geometrically distributed, with the mean $R_0+1$. The connected components are small. As contact tracing events (also for recursive tracing) cannot jump from one connected component to another one, correlations are strongly localized, and in that, we expect the marginal quantities to be appropriate to describe global properties of the the process. \par\medskip 
\begin{figure}
(a)\includegraphics[width=7cm]{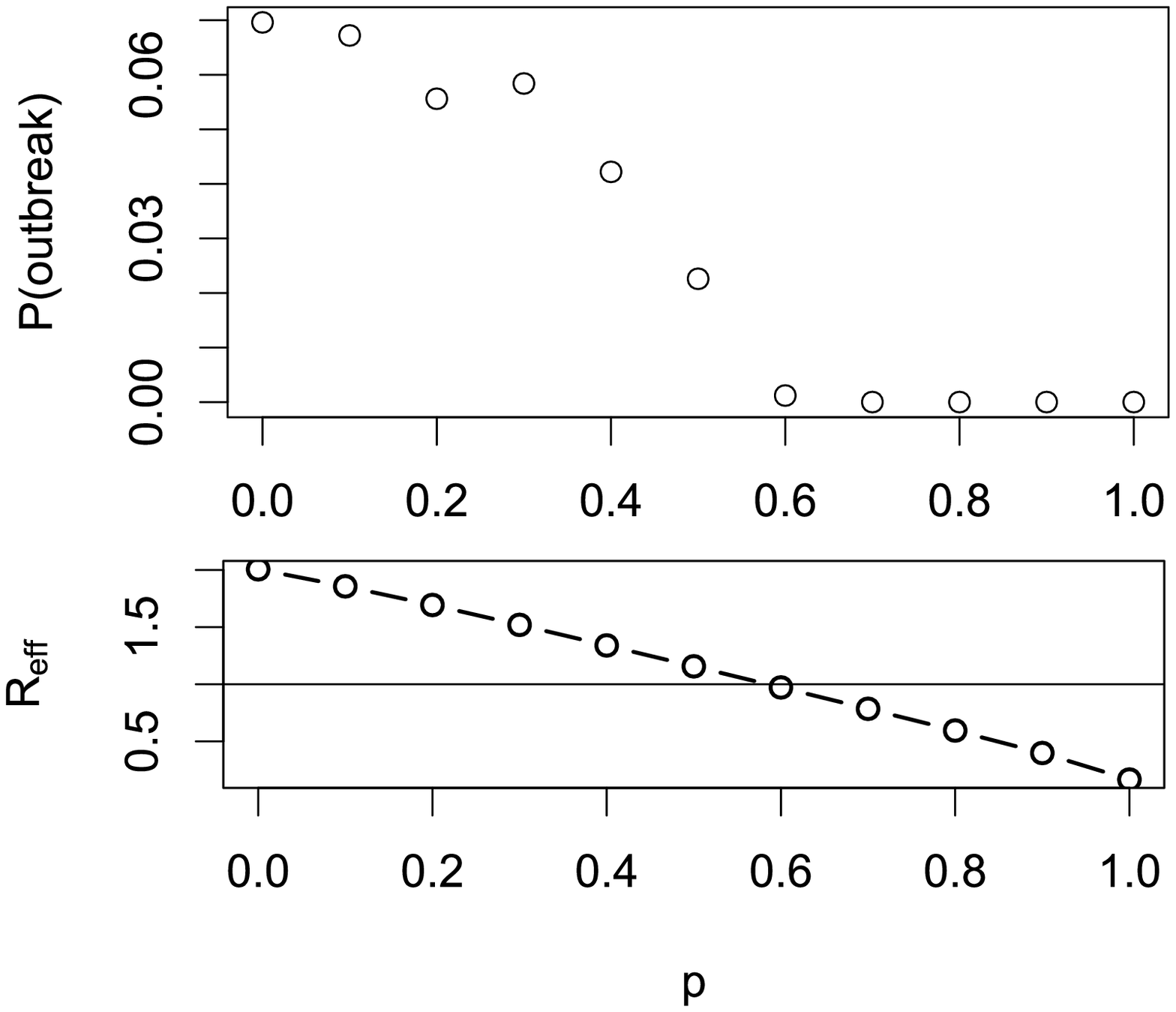}
\qquad (b)\includegraphics[width=7cm]{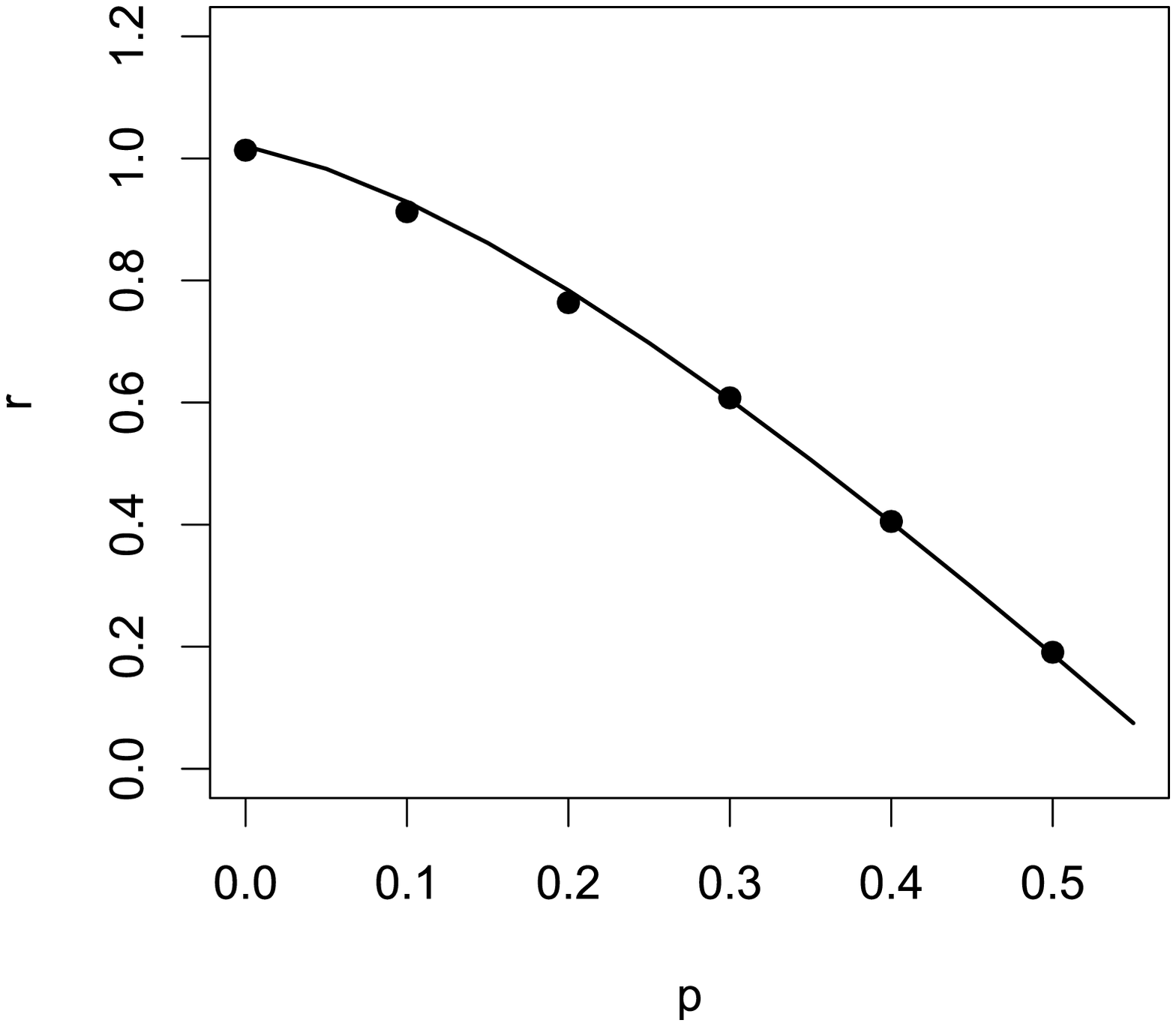}
\caption{
Comparison of simulations and theory for the full recursive branching-tracing process in dependence of $p$ (all three panels).  
(a) upper panel: probability of a major outbreak (fraction of 5000 simulations per parameter set that reach at least 8000 infeceteds and recovereds), lower panel: $R_{eff,4}$, horizontal line $R_{eff,4}=1$ 
(b) Growth rate (line: theory for $i=4$, bullets: simulated growth rate, averaged over 100 runs that are non-extinct).
Parameters: $\beta_0=0$, $\beta_1=2$, $\lambda=0.1$, $\mu=\sigma=0.5$, $p$ as indicated.
}
\label{compareTheoSimul}
\end{figure}
Simulations confirm that heuristic argument (Fig.~\ref{compareTheoSimul}), though this argument is -- particularly in case of extinction -- suspect: If the process goes extinct, we do not have many individuals available, s.t.\ even local correlations might be decisive.

\section{Effect of dispersion on CT}

We use our theory to investigate the impact of super-spreader events on 
the effect of CT. Thereto, we first consider the situation that all rates are constant, and in a second step, we aim to understand the dependency on 
incubation- and latency period. Last we inspect the results for a scenario that resembles the SARS-CoV-2 infection. \par\medskip 

In figure~\ref{R0Scan} we investigate what happens if all rates are constant (they do not depend on the age-since-infection). We recall that in case of $\lambda$ large implies that the system behaves as only deterministic contacts are present, while for $\lambda$ small super-spreader events dominate. In order to find the impact of super-spreader events, we particularly compare the lower and upper margin of the graphs / the direction of the curves. If the curves (that indicate lines of constant $R_{eff}$) tend from the left (lower margin) to the right (upper margin), then CT is more efficient with super-spreader events, and less efficient in the reverse case.\\ 
First of all, we find first of all that $R_{eff}$ changes in a clear interval for the frequency of super-spreader events. For the parameters chosen, that interval is approximately given by $\lambda \in [5,0.01]$. This observation is in line with the convergent results for $\lambda\rightarrow 0$ and $\lambda\rightarrow\infty$ that we worked out above. 
The observation, however, is stronger than convergence only. We can understand that the stripe for $\lambda$ were $R_{eff}$ changes is connected with timing: The infectious period of an individual is about $1/(\mu+\sigma)$ ($\approx 1$ in the simulation). If $\lambda$ is distinctively larger than $\mu+\sigma$, the average individual will have several (small) ``super-spreader'' events. As the events are frequent and small, they are indistinguishable from deterministic contact events, and $R_{eff}$ does not change if $\lambda$ is larger  than (about) $5(\mu+\sigma)$. In turn, if $\lambda$ is much smaller than $\mu+\sigma$, only very few individuals are hit by super-spreader events. These events are large indeed. We will find below that the effect of contact tracing does not change in case of large or very large super-spreader events. Therefore, $R_{eff}$ hardly does depend on $\lambda$ if either $\lambda\ll\mu+\sigma$ or $\lambda\gg\mu+\sigma$.\par\medskip 
\begin{figure}[htbp]
	\begin{center}
		\includegraphics[width=\textwidth]{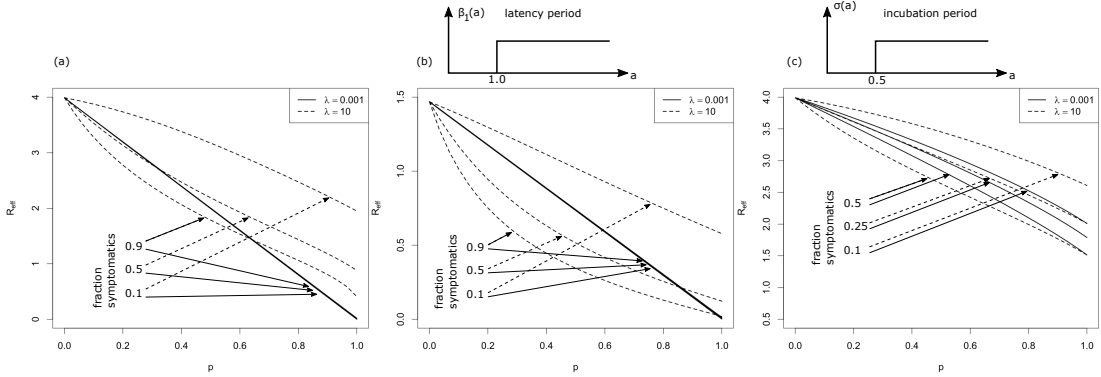}
	\end{center}
	\caption{$R_{eff}$ for $\lambda=10$ (dashed curves) and $\lambda=0.01$ (solid lines) over the tracing probability $p$ in case that the fraction of symptomatics is $0.1$, $0.5$ and $0.9$. Note that in case of $\lambda=0.001$ and panel (a), (b), the three lines are on top of each other. (a) All rates are constant, $\beta_0=0$, $\beta_1=4$, $\mu+\sigma=1$, and the fraction of symptomatics $\sigma/(\mu+\sigma)$ as indicated. (b) All rates but $\beta_1$ are as in (a), where $\beta_1=0$ for $a<1$, and $\beta_1=4$ for $a\geq 1$. (c) All rates but $\mu$ and $\sigma$ are as in (a); $\mu(a)+\sigma(a)\equiv 1$, $\sigma(a)=0$ for $a<0.5$, $\sigma(a)>0$ and constant for $a\geq 0.5$. }\label{sim8figure}
\end{figure}

\begin{figure}[h!]
	\begin{center}
		\includegraphics[width=\textwidth]{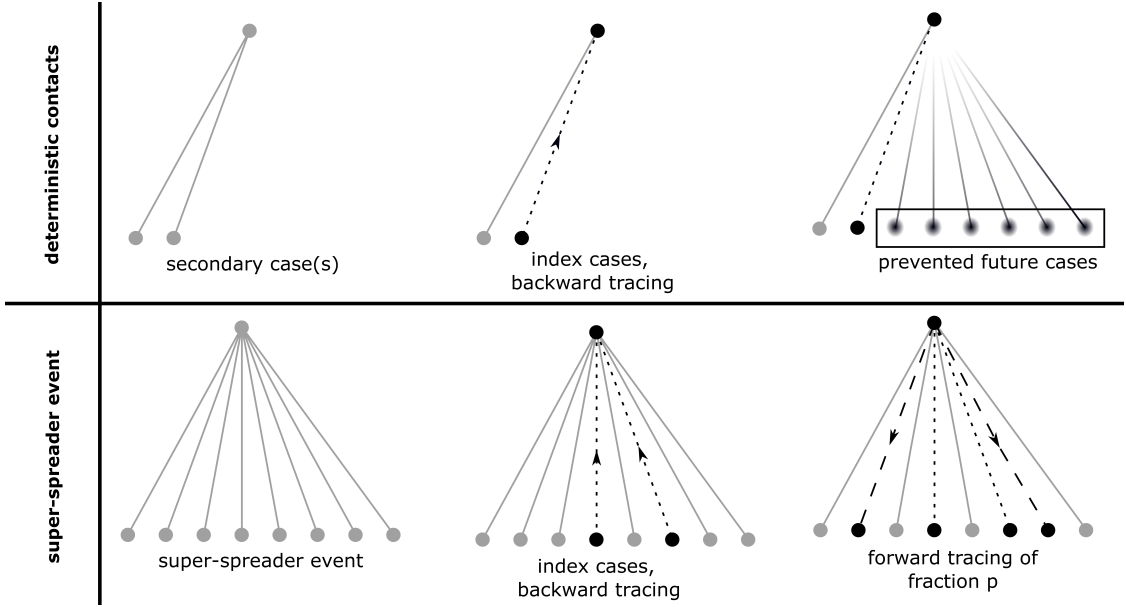}
	\end{center}
	\caption{Different mechanisms for super-spreader events and deterministic contacts: In super-spreader events, a fraction $p$ of infectees is {\it traced}. In deterministic contacts, a fraction of future infections is {\it prevented}; this fraction depends on the timing and can be larger than $p$.}\label{mechansims}
\end{figure}

Next we discuss the fact that CT seems always to perform better in the presence of super-spreader events if the fraction of symptomatics is small, while the effect of super-spreader events depends on the tracing probability if we have a high fraction of symptomatic cases. Thereto we compare $R_{eff}$ for small $\lambda$ ($=0.001$) with large $\lambda$ ($=10$), and for different fraction of symptomatic cases (Fig.~\ref{sim8figure}, panel (a)). The first fact to be noticed is that $R_{eff}$ is approximately linearly decreasing in $p$ for small $\lambda$, while it is non-linear if $\lambda$ is large. The second striking point is that $R_{eff}$ hardly depends on the fraction of symptomatics for $\lambda$ small, while for $\lambda$ large, $R_{eff}$ heavily depends on this fraction. CT is based on different mechanisms for these two cases (Fig.~\ref{mechansims}).\\ 
In case of $\lambda$ small, we have super-spreader events. As many infectees are produced synchronously, even in case a small fraction of symptomatics and $p$ is small, there will be very soon after the event sufficiently many index cases discovered, s.t.\ the super-spreader his/herself is detected by means of backward tracing. This fact explains the independence of $R_{eff}$ of the fraction of symptomatics. Once the super-spreader is known, a fraction $p$ of infectees are quarantined due to forward tracing. Therefore, the effect of CT  nicely fits a linear curve. \\
In case of deterministic contact events, a focal individual produces infectees, one after the other. Eventually an infectee will becomes symptomatic, and the focal individual can be traced. The higher the fraction of symptomatics, the earlier the focal individual will be detected by backward tracing.  Forward tracing might reveal more infectees. However, even more importantly, further infections are prevented. While in super-spreader events, a fraction $p$ of infectees are removed, in deterministic contact events -- depending on the timing -- the fraction of prevented cases might be much larger than $p$. As the timing crucially depends on the fraction of symptomatics in that case, also the effect depends on this fraction.
\\
Therefore, if $p$ is small and the fraction of symptomatic cases is high, CT can be more efficient in deterministic contacts events. If $p$ is large, then CT will be more efficient in case of super-spreader events. Realistic parameter settings are not that extreme. However, also in that case, CT will act in super-spreader events rather in detection of already infected persons, while in deterministic contacts it rather acts in preventing further infections by the focal individual. \\
In case of constant parameters and $\mu=0$, we have (see above) that the critical tracing probability coincides for the scenario with and without super-spreaders. In this special case, CT is more efficient without super-spreader events if $R_{eff}>1$, and less efficient for $R_{eff}<1$. If we also have asymptomatic cases, the effective reproduction number that separates the regions were CT is more/less efficient for super-spreader events increases.\par\medskip

Last we consider the effect of incubation and latent period. In Fig.~\ref{sim8figure}, panel (b), the effect of the latent period is presented. The figure does change quantitatively but not qualitatively. In case of a latency  period, CT is more efficient, as the tracing process is better able to catch up with the infectious process. Otherwise, the interpretation we obtained for the first case does not change.\\
That is different if we consider an increased incubation period. We adapted the model with constant rates s.t.\ $\mu(a)+\sigma(a)$ (the total recovery rate) still is constant, but that no individual shows symptoms in a first time interval. Therefore, there are less symptomatic cases and hence CT is less efficient. However, the most striking point is the fact that the curves for super-spreader events split up and become qualitatively similar to the curves for deterministic contacts. The basic mechanism for CT in case of super-spreader events changes. The infectees of the event cannot be removed early (as in the scenario without incubation period), and start to produce further infecteds until the super-spreader is detected.  In case of a long incubation period, a strong component in the effect of CT is the prevention of 
 further infecteds. That is, the mechanism moves from eliminating already infected persons to preventing further infections.

\begin{figure}
	\begin{center}
		\includegraphics[width=8cm]{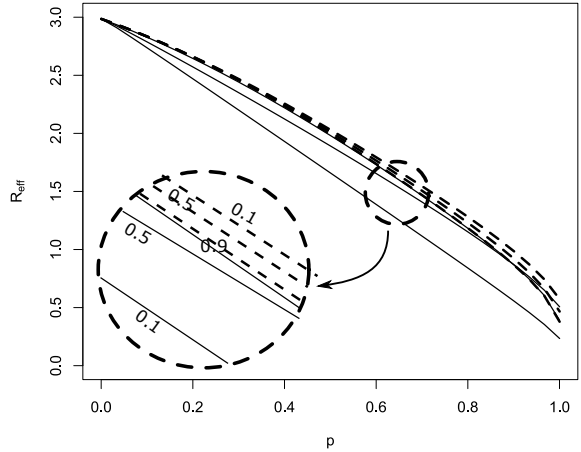}
	\end{center}
	\caption{For parameters resembling that for SRAS-CoV-2, we consider the impact of super-spreader events. $\beta_0(a)=\tau\beta(a)$, and $\beta_1(a)=(1-\tau)\beta(a)$.  Dashed: $\lambda=10$, solid lines $\lambda=0.001$. 
	$\tau$: as indicated in the inlay, symmetric from outside to the central curves, $\tau=0.1$, $\tau=0.5$, and $\tau=0.9$.}\label{covid}
\end{figure}

Let us turn to a real-world example: SARS-CoV-2. We use parameters that are derived using medical information about that infection (for details see~\cite{Pollmann2020}, SI, where we use here $R_0=3$ to calibrate). These parameters incorporate no information about supers-spreader events; we have only an idea about $\tilde \beta(a):=\beta_0(a)+\beta_1(a)$. Therefore we scan different possible ratios of $\beta_0$ and $\beta_1$, as well as different values for $\lambda$. We find in Fig.~\ref{covid} that $R_{eff}$ rather resembles the deterministic mechanism than the super-spreader mechanism for CT. And indeed, for SARS-CoV-2 the incubation period is about 2 days larger than the latency period. Due to this fact, CT is rather based on the prevention of further cases, and so super-spreader mechanisms do only change the efficiency of CT in a limited way.

\section{Discussion}

We considered a branching-type model for CT, where we introduced random contacts to model super-spreader events. The method developed in~\cite{mueller2000} was carried over to analyze the present model. Thereto, we introduced forward- and  backward tracing. The results for these sub-processes are interesting in itself. Particularly neither backward- nor forward tracing alone is able to adequately handle super-spreader events. In extreme cases, backward tracing alone has no effect at all. This observation has technical consequences (the combination of backward- and forward to full tracing requires more careful attention than in most studies based on~\cite{mueller2000} published up to now), but also practical consequences~\cite{Bradshaw2020}. Particularly, it is necessary to choose the tracing interval (the interval for which the contacts are identified) long enough to also detect the infector.\\
The second interesting finding of the present study is a difference in the mechanism of CT in deterministic- and super-spreader events. While in super-spreader events CT is based on a rapid detection of a fraction $p$ of infectees (and in that $R_{eff}$ decreases appropriately linearly while the fraction of symptomatic individuals only has a minor influence), in deterministic, non-synchronous contacts, CT is rather based on the prevention of cases. If the tracing probability is small and the fraction of symptomatics (index-cases) is large, then CT becomes less effective in case of super-spreader events. A large incubation period (distinctively larger than the latency period) relativize the differences - the ``super-spreader mechanism'' becomes less important, and the results resemble more the situation with deterministic contacts. Particularly, in case of SARS-CoV-2, we don't recognize a huge effect.\\

The present paper deepens our understanding of CT in the presence of super-spreader events. We are able to better understand the different mechanisms by which CT controls an infection with/without super-spreader events. This insight, in turn, allows to characterize situation where the effect of CT is not increased but decreased.\\ 
In any case, we need to emphasize that the present  model only is one of many possible ways to address superspreading. Another approach is based on a heterogeneous contact graph. Also in that case, there are indications that a higher variance (superspreading) decreases the efficiency of contact tracing (see Okolie et al.~\cite{Okolie2020}). However, in the present study as well as in the paper by Okolie et al.~\cite{Okolie2020}, we assume that  only infector/infectee links can be traced, and all links independently. It might be more realistic that also direct contacts happen between infectees in a superspreader events, or that the complete group of infectees is detected. Both cases violate assumption that contacts/individuals are detected independently of each other.  In that, the present approach might underestimate the efficacy of CT. Further investigations are necessary to analyze the different reasons for super-spreading, and the effect of super-spreading on the efficacy of CT.
\par\bigskip 

{\bf Acknowledgements} This research is supported by a grant from the Deutsche Forschungsgemeinschaft (DFG) through TUM International Graduate School of Science and Engineering
(IGSSE), GSC 81, within the project GENOMIE QADOP (JM).

\bibliographystyle{abbrvurl}
\bibliography{ctBib}

\end{document}